\documentclass[12pt]{article}
\newlength{\abstractwidth}
\usepackage{epsf}

\renewcommand{\thefootnote}{\fnsymbol{footnote}}
\renewcommand{\thanks}[1]{\footnote{#1}}
\newcommand{\starttext}{
\setcounter{footnote}{0}
\renewcommand{\thefootnote}{\arabic{footnote}}}

\newcommand{\bea}{\begin{eqnarray}}
\newcommand{\eea}{\end{eqnarray}}
\newcommand{\ee}{\end{equation}}
\newcommand{\be}{\begin{equation}}

\def\cA{{\cal A}}
\def\cB{{\cal B}}
\def\cC{{\cal C}}

\def\cF{{\cal F}}
\def\cG{{\cal G}}
\def\cH{{\cal H}}
\def\cI{{\cal I}}
\def\cJ{{\cal J}}
\def\cK{{\cal K}}

\def\cN{{\cal N}}
\def\cO{{\cal O}}

\def\cT{{\cal T}}
\def\cU{{\cal U}}

\def\bR{{\bf R}}

\def\Re{{\rm Re}}
\def\Im{{\rm Im}}

\def\Tr{{\rm Tr}}
\def\det{{\rm det}}

\def\half{ {1\over 2}}
\def\p{\partial}

\def\tet{\vartheta}

\def\a{\alpha}
\def\b{\beta}
\def\g{\gamma}

\def\ep{\varepsilon}
\def\o{\omega}

\def\G{\Gamma}

\def\ch{{\rm ch \, }}
\def\sh{{\rm sh \, }}

\def\cotg{{\rm cotg}}

\def\no{\nonumber}

\begin{document}
\starttext
\setcounter{footnote}{0}

\begin{flushright}
UCLA/07/TEP/11 \\
8 May 2007
\end{flushright}

\bigskip

\begin{center}

{\Large \bf Gravity duals of half-BPS Wilson loops}

\vskip .7in

{\large  Eric D'Hoker,  John Estes and  Michael Gutperle}

\vskip .2in

 \sl Department of Physics and Astronomy \\
\sl University of California, Los Angeles, CA 90095, USA

\end{center}

\vskip .5in

\begin{abstract}
We explicitly construct the fully back-reacted half-BPS solutions in Type IIB 
supergravity which are dual to Wilson loops with 16 supersymmetries in 
$\cN=4$ super Yang-Mills. In a first part, we use the methods of a companion 
paper to derive the exact general solution of the half-BPS equations on the 
space $AdS_2 \times S^2 \times S^4 \times \Sigma$, with isometry group
$SO(2,1)\times SO(3) \times SO(5)$ in terms of two locally harmonic functions 
on a Riemann surface $\Sigma$ with boundary. These solutions,
generally, have varying dilaton and axion, and non-vanishing 3-form fluxes.
In a second part, we  impose regularity and topology conditions. 
These non-singular solutions may be parametrized by a genus $g \geq 0$ 
hyperelliptic surface $\Sigma$, all of whose branch points lie on the real line. 
Each genus $g$ solution has only a single asymptotic $AdS_5 \times S^5$
region, but exhibits $g$ homology 3-spheres, and an extra $g$ homology
5-spheres, carrying respectively RR 3-form and RR 5-form charges.
For genus 0, we recover $AdS_5 \times S^5$ with 3 free parameters,   
while for genus $g \geq 1$, the solution has $2g+5$ free parameters.
The genus 1 case is studied in detail. Numerical analysis is 
used to show that the solutions are regular throughout the $g=1$ parameter space. 
Collapse of a branch cut on $\Sigma$ subtending either a homology 3-sphere or a 
homology 5-sphere is non-singular and yields the genus $g-1$ solution. 
This behavior is precisely expected of
a proper dual to a Wilson loop in gauge theory.

\end{abstract}

\newpage

\baselineskip=16pt
\setcounter{footnote}{0}

\section{Introduction}
\setcounter{equation}{0}
\label{one}

In the AdS/CFT correspondence, 
\cite{Maldacena:1997re, Gubser:1998bc, Witten:1998qj} 
(for reviews, see \cite{D'Hoker:2002aw, Aharony:1999ti}) a prominent role  
is played by objects which are protected by supersymmety. 
Non-renormalization theorems make comparison of weak and strong coupling 
calculations possible and BPS equations often provide an easier way to obtain 
exact solutions of  the equations of motion.

\smallskip

The first example of such objects is provided by local gauge invariant  chiral primary 
operators ${\cal O}_J$ in $\cN=4$ super Yang-Mills theory (SYM) with gauge group $SU(N)$, 
where $J$ denotes a particular $U(1)$ charge of the $SO(6)$ R-symmetry.  
Supersymmetry protects the conformal dimension $\Delta$ of these operators 
against quantum corrections, so that we have the exact relation $\Delta = J$.
For $\Delta \ll N$, these operators are dual to small fluctuations 
of supergravity modes in the $AdS_5\times S^5$ bulk \cite{Witten:1998qj}. 
For $\Delta \sim N$, they can be associated with giant gravitons  
(probe branes on $S^{5}$ or $AdS_5$) \cite{Brodie:2000yz,Hashimoto:2000zp,Grisaru:2000zn}. For $\Delta\sim N^2$, the fully back-reacted geometries preserving $SO(4)\times SO(4)$ 
as well as 16 of the 32 supersymmetry were found in \cite{Lin:2004nb} and referred to
as ``bubbling AdS".  All regular solutions are parameterized by the ``coloring" of $\bR^2$  
in black and white regions. 

\smallskip

In two recent papers \cite{EDJEMG1,EDJEMG2}, the present authors constructed 
general half-BPS solutions with $SO(2,3) \times SO(3)\times SO(3)$ symmetry
in Type IIB supergravity. 
The solutions are given by a warping of $AdS_4\times S^2\times S^2$ over a 
two-dimensional surface $\Sigma$, and have varying dilaton and axion, as well as 
non-vanishing NSNS  and RR 3-form fluxes. They generalize the 
non-supersymmetric \cite{Bak:2003jk} and ${\cal N}=1$ supersymmetric 
\cite{D'Hoker:2006uu} Janus solutions. 
These solutions are holographic duals of (generalized) interface SYM theories \cite{D'Hoker:2006uv,Clark:2004sb}.

\smallskip

The solutions were found by solving the  BPS equations of Type IIB supergravity for the 
most general  $SO(2,3) \times SO(3)\times SO(3)$ symmetric Ansatz on the manifold
$AdS_4\times S^2\times S^2 \times \Sigma $ (see also \cite{Gomis:2006cu}). 
The solutions are parametrized by two harmonic functions $h_1,h_2$ on a 
genus $g$ hyperelliptic Riemann surface $\Sigma$ with boundary, with all the branch 
cuts restricted to lie on the real axis. The regularity of the solutions imposes various 
conditions on the harmonic functions. We refer the reader to the papers  \cite{EDJEMG1,EDJEMG2}  for details. The choice of $g+1$ branch cuts along the real axis 
is the one-dimensional analog of the coloring of the ``bubbling AdS" solution of \cite{Lin:2004nb}.

\smallskip

Another important class of operators consists of  Wilson loops corresponding to the 
holonomies of gauge fields along (closed) contours. The AdS dual of a Wilson loop 
operator in the fundamental representation was identified in 
\cite{Maldacena:1998im,Rey:1998ik}, with a fundamental string world sheet in the bulk
of $AdS$ which ends on the contour of the Wilson loop on the boundary of $AdS$. 
In particular, we will be interested in half-BPS Wilson loop operators.  Proposals for the 
AdS-dual description of half-BPS Wilson loop operators in higher dimensional 
representations of $SU(N)$ have been made in \cite{Drukker:2005kx,Gomis:2006sb,Gomis:2006im,Yamaguchi:2006te, Hartnoll:2006hr}

\smallskip

In this paper, the fully back-reacted supergravity solutions corresponding to half-BPS 
Wilson loops will be derived and explicit formulas for the solutions will be presented. 
We shall follow closely the methods developed for the half-BPS Type IIB interface 
solutions in \cite{EDJEMG1,EDJEMG2}. In particular, we shall solve the BPS equations 
for our Ansatz explicitly in terms of two harmonic functions $h_{1},h_{2}$ defined on a 
two-dimensional Riemann surface $\Sigma$. Indeed, many formulas in the present 
paper will be   very similar or identical to the ones in \cite{EDJEMG1,EDJEMG2}. 
There are, however,  subtle and important differences between  the solutions.   
Furthermore, while it is possible to formally relate the solutions on 
$AdS_4\times S^2\times S^2\times \Sigma$ of  \cite{EDJEMG1,EDJEMG2} to the 
solutions 
on $AdS_2\times S^2\times S^4\times \Sigma$ given in the present paper by an analytic continuation, it is not a priori guaranteed that both sets of solutions will preserve the same number of supersymmetries or even that regular solutions will be mapped to regular
solutions. A supergravity description of half-BPS Wilson loops has already 
been given in \cite{Lunin:2006xr} (See also \cite{Yamaguchi:2006te} for an earlier attempt 
at a solution), using the Killing tensor  methods of 
\cite{Lin:2004nb,Gauntlett:2002nw,Gillard:2004xq}. The solution found in \cite{Lunin:2006xr}
was parameterized by  a harmonic function but some quantities were only implicitly given in 
terms of the harmonic function.

\begin{figure}[tbph]
\begin{center}
\epsfxsize=5.3in
\epsfysize=3.4in
\epsffile{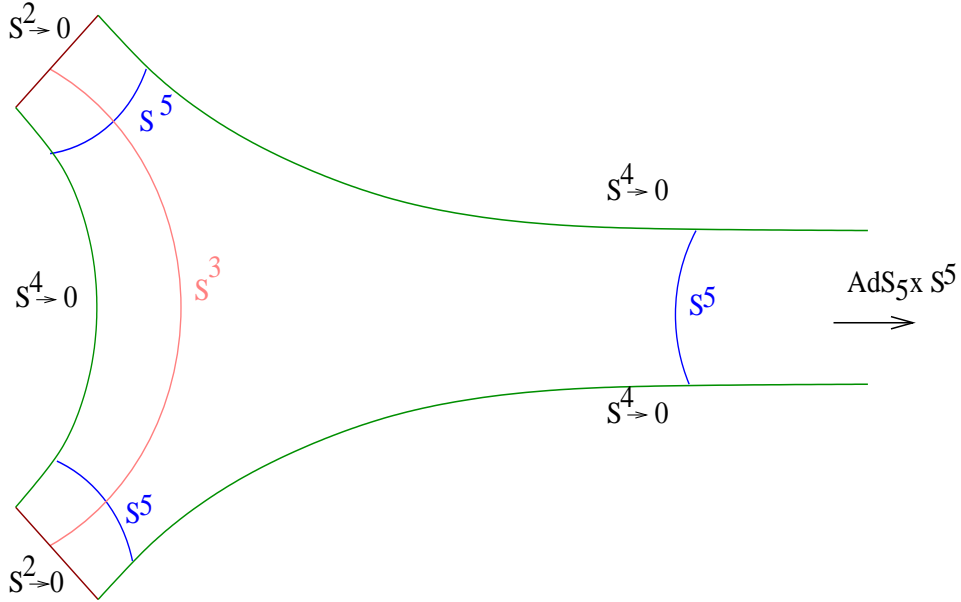}
\label{figure1}
\caption{The genus 1 solution with homology 3-spheres and 5-spheres.}
\end{center}
\end{figure}

 A summary of the key properties of our AdS solutions is as follows. The 
$SO(2,1) \times SO(3) \times SO(5)$-invariant metrics on  
$AdS_2\times S^2\times S^4\times \Sigma$ are parametrized  by 
\bea
ds^2 = f_1 ^2 ds^2 _{AdS_2} + f_2 ^2 ds^2 _{S^2} + f_4^2 ds^2 _{S^4}
+ ds^2 _\Sigma
\eea
The metrics $ds^2 _{AdS_2}$, $ds^2 _{S^2}$, and $ds^2 _{S^4}$
correspond to unit radii.
The metric $ds^2 _\Sigma$ is positive, and may be expressed as 
$ds^2 _\Sigma = 4 \rho ^2 |dw|^2$ in terms of local complex coordinates $w,\bar w$ 
on the Riemann surface $\Sigma$ with boundary $\p \Sigma$.
The dilaton $\phi$, and $\rho, f_1,f_2,f_4$ are real functions on $\Sigma$. 
All half-BPS solutions may be expressed in terms of two real harmonic functions 
$h_1$ and $h_2$ on $\Sigma$. The dilaton and metric functions for these
solutions are given by the following relations, 
\bea
\label{dil0}
e^{4 \phi} = - \, 
{2 h_1 h_2 |\p_w h_2|^2 - h_2 ^2 (\p_w h_1 \p_{\bar w} h_2 +
\p_w h_2 \p_{\bar w} h_1) 
\over 
2 h_1 h_2 |\p_w h_1|^2 - h_1 ^2 (\p_w h_1 \p_{\bar w} h_2 +
\p_w h_2 \p_{\bar w} h_1)}
\eea
as well as
\bea
f_2^2 f_4^2 & = & 4 h_2 ^2 \, e^{- 2 \phi}
\no \\
f_1^2 f_4^2 & = & 4 h_1^2 \, e^{+ 2 \phi}
\eea
Explicit expressions for the solutions of $\rho$, $f_1$ are given in (\ref{rho})
and (\ref{radii-}). (The Ansatz for the antisymmetric tensor fields, will be  given 
in (\ref{AST1}), (\ref{AST2}), (\ref{AST3}), and their solutions obtained in
section 7.6.) 

\smallskip

For all regular solutions,  the functions $f_1, f_2, f_4$, and $e^{\pm 2 \phi}$
are non-vanishing inside $\Sigma$.  The boundary $\p \Sigma$
has either the $S^2$ or the $S^4$ shrink to 0 radius, so that either $f_2$ or $f_4$ vanishes
respectively, and $h_2=0$ throughout  $\p \Sigma$. Since $f_1\not= 0$ on $\p \Sigma$,
we also have $h_1=0$ on segments of $\p \Sigma$ if and only if the $S^4$
shrinks to zero on that segment. All solutions have a single asymptotic 
$AdS_5 \times S^5$ region. A schematic picture for the case where $\Sigma$
is a genus 1 surface is given in Figure 1. The presence of a non-trivial 
homology 3-sphere indicates the presence of a non-vanishing RR 3-form charge
as will be computed in section 10.5.
 
The plan of this paper is as follows. 
In section \ref{two}, we review half-BPS Wilson loops in the context of AdS/CFT. 
In section \ref{three}, we present our Ansatz as $AdS_{2}\times S^2 \times S^4$ 
warped over a two dimensional space $\Sigma$. The general ten-dimensional Killing 
spinor is decomposed with respect to the $AdS_{2}\times S^2 \times S^4$  factors. 
In section \ref{four}, the BPS equations are reduced by utilizing the Killing spinors. 
In section \ref{five}, reality conditions are imposed on the solution and the BPS equations 
are reduced to equations on a two dimensional complex spinor. 
In section \ref{six},  these equations are reduced to an integrable system. 
In section \ref{seven}, the integrable system is mapped to a first order system which 
can be solved in terms of two harmonic functions. The explicit form of the metric factor, 
dilaton and three form fluxes in terms of the harmonic functions are given. 
In section \ref{eight}, it is shown that the only solution with a constant dilaton is  
$AdS_{5}\times S^{5}$. 
In section \ref{nine}, the conditions for obtaining regular solutions are derived and 
the boundary conditions on the harmonic functions are derived. 
In section \ref{ten}, a general class of regular solutions is constructed with a genus $g$ hyperelliptic surface. 
In section \ref{eleven}, the genus 1 case is discussed in detail and all the quantities are  explicitely expressed in terms of elliptic functions. 
In section 12, we study the collapse of a branch cut on $\Sigma$ between 
consecutive branch points, and show that the genus $g$ solution collapses
to a regular genus $g-1$ solution.
In three appendices we give our conventions for the Clifford algebra, Killing spinors,
Bianchi identities and equations of motion.

\newpage

\section{Wilson loops in gauge theory and supergravity}
\label{two}
\setcounter{equation}{0}

The Wilson loop operator is an important gauge invariant observable in gauge theories. 
For ${\cal N}=4$ super Yang-Mills theories, the appropriate operator is defined as 
\cite{Maldacena:1998im,Rey:1998ik,Drukker:1999zq}
\be
W_{R}(C)= 
\Tr_{R}\; P \exp \left ( i \int_{C}d\tau ( A_{\mu} \dot x^{\mu}+ \phi_{I}\dot y^{I}) \right )
\ee
Here, $\Tr_{R}$ labels the trace over an arbitrary representation $R$ of $SU(N)$ 
and $(x^{\mu}(\tau),y^{I}(\tau))$ parameterizes a path in $R^{1,3}\times R^{6}$ coupling
the curve $C$ to the gauge field and the six adjoint scalars of ${\cal N}=4$ SYM. 
The Wilson loop operator can preserve some supersymmetry. In particular, 
it was shown in \cite{Drukker:1999zq} that the preservation of eight Poincar\'e supersymmetries restricts 
the path in $R^{1,9}$ to be null, i.e. $\dot x^{2}+\dot y^{2}=0$. The preservation of eight superconformal symmetries furthermore fixes the trajectory on $R^{1,3}$ to be a timelike 
line, i.e. $x^{0}=\tau, x^{i}=0$ and the trajectory on $R^{6}$ to be given by $\dot y^{I}=n^{I}$, where $n^{I}$ is a unit vector in $R^{6}$. Hence the half-BPS Wilson loop operator in the representation $R$ becomes
 \be
W_{R}(C)= \Tr_{R}\; P \exp \left ( i \int_{C}d\tau (A_{0}+ n_{I}\phi^{I}) \right )
\ee

\subsection{Symmetries of the half-BPS Wilson loop}

In the following, we review the derivation of the supergroup preserved by the half-BPS 
Wilson loop. The choice of the unit vector  $n^{I}$ breaks the $SO(6)$ R-symmetry to $SO(5)$. 
The superconformal symmetry $SO(4,2)$ is broken by the timelike straight line as follows. 
The condition $x^{i}=0, ~i=1,2,3$ is left invariant by $SO(3)$ spatial rotations. 
The condition $\dot  x^{0}=1$ is invariant under time translations, dilations and special 
conformal transformations which together generate $ SO(2,1)$. 
Hence the residual bosonic symmetry is
\be
 SO(2,1)\times SO(3)\times SO(5)
 \ee
The sixteen unbroken supersymmetries transform under the $(4,4)$ of the bosonic symmetry group and form a supergroup $OSP(4^{*}|4) $. 

\smallskip

For the supergravity description of the half-BPS Wilson loops, we seek a general Ansatz 
in Type IIB supergravity with the above symmetry.
The factor $SO(2,1)$ requires the geometry to contain $AdS_2$,
the factor $SO(3)$ requires $S^2$, and the factor $SO(5)$ requires
$S^4$. Two dimensions remain undetermined by the symmetries alone, so
that the most general space of interest to us will be of the form,
\bea
\label{eqwarpa}
AdS_2 \times S^2 \times S^4 \times   \Sigma
\eea
where $\Sigma$ stands for the two-dimensional parameter space,
over which the above products are warped.

\subsection{Geometry of fluxes}

The holographic dual description of a (probe) Wilson loop operator in the fundamental 
representation of $SU(N)$ is given by a string worldsheet in the $AdS_{5}\times S^{5}$ 
bulk which ends on the contour $C$ of the Wilson loop on the boundary of $AdS_5 \times S^5$.  
The holographic description of half-BPS Wilson loop operators in higher dimensional 
representations of $SU(N)$ was developed by several authors 
\cite{Drukker:2005kx,Gomis:2006sb,Gomis:2006im,Yamaguchi:2006te}. 
It appears that there are two equivalent descriptions in terms of D-branes with 
fundamental string charge.

\smallskip

In the first proposal \cite{Gomis:2006sb,Gomis:2006im,Yamaguchi:2006te},
a Wilson loop in the $k$-th symmetric tensor representation (which is labeled by a 
Young-tableau of a row of $k$ boxes) is given by a D3 brane with $AdS_{2}\times S^{2}$ 
worldvolume with  $k$ units of fundamental string charge dissolved on the $AdS_{2}$. 
A  general Young-tableau with $p$ rows with $n_{i}, ~i=1,\cdots,p$ is given by an 
array of $D{3}$ branes with $n_{i}$ units of fundamental string charge dissolved on 
the $i$-th D3-brane.

\smallskip

In the second proposal \cite{Drukker:2005kx,Gomis:2006sb},
a Wilson loop in the $k$-th anti-symmetric tensor representation (which is labeled by a Young-tableau of a column  of $k$ boxes) is given by a D5 brane with worldvolume $AdS_{2} \times S^{4}$ where $k$ units of fundamental string charge are dissolved on the $AdS_{2 }$ worldvolume. Similarly to the first proposal a general representation with $q$ columns with $m_{i},~ i=1,\cdots,q$ boxes is given by an array of D5-branes with $m_{i}$ units of fundamental string charge disolved on the $i$-th D5-brane. 

\smallskip

In both descriptions the D-branes are effectively treated as probes and the back-reaction 
on the geometry is neglected. The probe brane description of the Wilson loops is, however, 
very useful to determine the correct ansatz for the fluxes and scalars. 
The Born-Infeld form of the action for a Dp-brane is given by
\be
S= \tau_{p}\int d^{p+1}\sigma ~ e^{-\Phi} \sqrt{- \det(g + B + 2\pi\alpha' F)}
+\mu_{p}\int C \wedge e^{2\pi \alpha' F} 
\ee
A non-vanishing fundamental string charge manifests itself as a non-zero electro-magnetic 
field strength $F_{\mu\nu}$ in the $AdS_{2}$ worldvolume direction.
 In the probe approximation the worlvolume electric field will be a source for other 
 supergravity fields. 
 
\smallskip

 For a probe  D3-brane with $AdS_{2}\times S^{2}$ worldvolume the Born-Infeld action 
 contains the following term linear in the electric field
\be
2\pi \alpha' \tau_{4} \int d^{4} \sigma\; e^{-\Phi} B_{\mu\nu} F^{\mu\nu} +2\pi \alpha' \mu_{4} \int F\wedge C_{2} +  \mu_{4} \int  C_{4}
\ee
  Consequently the electric field sources the NSNS two form potential $B_{2}$ in the $AdS_{2}$ directions and the RR two form potential $C_{2}$ in the $S^{2}$ direction.

\smallskip

 For a probe  D5-brane with $AdS_{2}\times S^{4}$ worldvolume the Born-Infeld action contains the following term linear in the electric field
\be
2\pi \alpha' \tau_{6} \int d^{6} \sigma\; e^{-\Phi} B_{\mu\nu} F^{\mu\nu} +\mu_{6} \int   C_{6} + 2\pi \alpha' \mu_{6} \int F\wedge C_{4} 
\ee
Consequently, the electric field sources the NSNS two form potential $B_{2}$ in the $AdS_{2}$ directions  and the RR six form $C_{6 }$ potential  along $AdS_{2}\times S^{4}$. By electromagnetic duality the six form potential is related to the RR two form potential $C_{2}$ which will be sourced along the $S^2$ direction. Note that the strength of the probe brane sources differ in the two cases the direction in which the NSNS and RR fluxes are sourced are the same.  
In addition the probe brane sources the dilaton but not the RR axion. 

\smallskip

In the rest of the paper  the completely back-reacted solution which is dual to half-BPS 
Wilson loops will be constructed.

\newpage

\section{The Ansatz}
\setcounter{equation}{0}
\label{three}

As discussed in section \ref{two}, the half-BPS Wilson loop preserves the bosonic 
symmetries $SO(2,1)\times SO(3)  \times SO(5)$  and is invariant under 16 supersymmetries. Our conventions for Type IIB supergravity will follow the ones used in
\cite{EDJEMG1,EDJEMG2} which coincide with those of \cite{Schwarz:1983qr}.
As a Type IIB supergravity geometry, $\Sigma$ will carry an orientation as well as a 
Riemannian metric, and is therefore a Riemann surface, generally non-compact, 
and with boundary. The symmetries determine the general form of the metric and 
the presence of the fundamental string charge and D5 branes dtermine the presence 
of NSNS and RR three form fluxes, leading to the following Ansatz for the supergravity fields.

\subsection{Ansatz for the Type IIB fields}

The appropriate supergravity ansatz for the metric is given by  (\ref{eqwarpa}) which is a warped product  of $AdS_2 \times S^2 \times S^4$ factors over a two dimensional surface  $\Sigma$. 
\bea
ds^2 = f_1 ^2 ds^2 _{AdS_2} + f_2 ^2 ds^2 _{S^2} + f_4^2 ds^2 _{S^4}
+ ds^2 _\Sigma
\eea
where $f_1,f_2,f_4$ and $ds^2 _\Sigma$ are real functions on $\Sigma$. We
introduce an orthonormal frame,
\bea
\label{frame1}
AdS_2 & \hskip 0.7in & e^\mu = f_1 \, \hat e^\mu \hskip 1in \mu =0,1
\no \\
S^2 \quad & & e^i = f_2 \, \hat e^i \hskip 1.08in i =2,3
\no \\
S^4 \quad && e^m = f_4 \, \hat e^m \hskip 0.9in m=4,5,6,7
\no \\
\Sigma \hskip 0.2in && e^a  \hskip 1.57in a=8,9
\eea
where $\hat e^\mu$, $\hat e^i$, $\hat e^m$,  and $e^a$ refer to orthonormal
frames for the spaces $AdS_2$,  $S^2$,  $S^4$,  and $\Sigma $ respectively.
In particular, we have\footnote{The convention of summation over repeated
indices will be used throughout whenever no confusion is expected to arise,
with the ranges of the various indices following the pattern of the frame in (\ref{frame1}). Complex frame indices on $\Sigma$ will often be used,
with the following conventions, $e^z = (e^8 + i e^9)/2$, $e^{\bar z} 
= (e^8 - i e^9)/2$, and the non-vanishing components of the metric
on $\Sigma$ are given by $\delta _{z \bar z} = \delta _{\bar z z}=2$.}
\bea
ds^2 _{AdS_2} = \eta _{\mu \nu} \hat e^\mu \otimes \hat e^\nu
& \hskip 1in &
ds^2 _{S^2} = \delta _{ij} \, \hat e^i \otimes \hat e^j
\no \\
ds^2 _{S^4} = \delta _{mn} \, \hat e^m \otimes \hat e^n 
& \hskip 1in &
ds^2 _\Sigma = \delta _{ab} \, e^a \otimes e^b \hskip .55in
\eea
where $\eta = {\rm diag} [- +]$. The  dilaton and axion fields  are represented by the 
1-forms $P$ and $Q$ which vary over $\Sigma$, 
and whose structure is  given as follows,
\bea
P& = & p_a e^a, \quad \quad 
Q  =  q_a e^a
\eea
while the anti-symmetric tensor forms $F_{(5)}$ is self dual and given by
\bea
F_{(5)} & = & \bigg( - e^{0123} \wedge \cF + e^{4567} \wedge *_{2} \cF \bigg)
\eea
In agreement with the symmetries and the probe analysis the three form field strength is  constructed from the unit volume form on $AdS_{2}$ and $S^{2}$,
\bea
G & = & e^{01} \wedge \cG +  i e^{23} \wedge \cH
\eea
where we have introduced the following 1-forms on $\Sigma$ to represent the reduced fields,
\bea
\cG  \equiv  g_a e^a & \hskip 1in & \cF  \equiv  f_a e^a
\no \\
\cH  \equiv  h_a e^a && *_{2} \cF \equiv \ep^a{}_b \, f_a e^b 
\hskip 0.7in \ep ^{89}=+1
\eea
Here, $f_a, q_a$ are real, while $g_a, h_a, p_a$ are complex.

\subsection{The general ten-dimensional Killing spinor}

The requirement that 16 supersymmetries remain preserved by the Ansatz puts severe 
restrictions on the supergravity fields, which result from enforcing the BPS equations,  
$\delta \lambda = \delta \psi _M=0$. Whenever the dilaton is subject to a non-trivial 
space-time variation, $\p_M \phi \not=0$, the dilatino BPS equation $\delta \lambda=0$ 
will allow for at most 16 independent supersymmetries $\ep$. Therefore, the gravitino 
BPS equation cannot impose any further restrictions on the number of supersymmetries,
but should instead simply give the space-time evolution of $\ep$.
Thus, at any fixed point in the parameter space $\Sigma$, $\ep$
must be a Killing spinor on each of the spheres $S^2$ and $S^4$,
as well as on $AdS_2$.  The Killing spinor equations on
$AdS_2 \times S^2   \times S^4 $ are given by
\bea
\label{KS}
\left ( \hat \nabla _\mu  - {1 \over 2} \eta_1\, \gamma _\mu \otimes I_2 \otimes I_4 \right )
\chi ^{\eta _1, \eta _2, \eta _3} _{\eta _0}
    & = & 0 \hskip 1in \mu =0,1
\no \\
\left ( \hat \nabla _i - {i \over 2} \eta_2 \, I_2 \otimes \gamma _i \otimes I_4 \right )
\chi ^{\eta _1, \eta _2, \eta _3} _{\eta _0}
    & = & 0 \hskip 1in i=2,3
\no \\
\left ( \hat \nabla _m - {i \over 2} \eta_3\,  I_2 \otimes I_2 \otimes \gamma _m \right )
\chi ^{\eta _1, \eta _2, \eta _3} _{\eta _0}
    & = & 0 \hskip 1in m=4,5,6,7
\eea
Here, $\hat \nabla _\mu$, $\hat \nabla _i$, and $\hat \nabla_m$,  are respectively
the covariant derivatives acting in the Dirac spinor representations for $AdS_2$, $S^2$,
and $S^4$, with respect to the canonical spin connections associated
with the frames  $\hat e^\mu$, $\hat e^i$ and $\hat e^m$.
The spinors $\chi ^{\eta _1, \eta _2, \eta _3} _{\eta _0}$ are 16-dimensonal.
The integrability conditions for each of these equations are automatically satisfied,
so that for fixed $\eta$, the number of independent (complex) Killing
spinors is respectively 2, 2 and 4 for the three equations. The indices
$\eta _0, \eta_1, \eta _2, \eta _3$  are independent and may take values $\pm 1$,
and therefore uniquely label a basis of the 16-dimensional spinor space.
The label $\eta _0$ arises because the $S^4$ equation is for a 4-component
spinor, whose solutions are labeled by the pair $(\eta _0, \eta _3)$.
Since the Killing spinor equation does not actually depend on $\eta_0$,
this index simply labels two linearly independent spinors for which
the reduced BPS equations are identical. Therefore, the index $\eta _0$
will be dropped, with the understanding that the solution space for
$\chi ^{\eta _1, \eta _2 , \eta _3}$ remains 16-dimensional.
(See \cite{EDJEMG1} for the discussion of the analogous issues for $AdS_4$.)

\smallskip

For any one of the chirality matrices $\gamma _{(s)}$, for $s=1,2,3$, the product
$\gamma _{(s)} \chi$ satisfies (\ref{KS}) with the opposite
value of $\eta _s$. We may therefore identify the corresponding spinors,
\bea
\label{chiralities}
\bigg( \gamma _{(1)} \otimes I_2 \otimes I_4 \bigg) \chi ^{\eta _1, \eta _2, \eta _3}
    & = & \chi ^{-\eta _1, \eta _2, \eta _3}
\no \\
\bigg(I_2 \otimes \gamma _{(2)} \otimes I_4 \bigg) \chi ^{\eta _1, \eta _2, \eta _3}
    & = & \chi ^{\eta _1, -\eta _2, \eta _3}
\no \\
\bigg( I_2 \otimes I_2 \otimes \gamma _{(3)} \bigg) \chi ^{\eta _1, \eta _2, \eta _3}
& = & \chi ^{\eta _1, \eta _2, -\eta _3}
\eea
To examine the Killing spinor properties, we begin by decomposing the 32
component (complex) spinor $\ep$ onto the $\Sigma$-independent basis of spinors
$\chi ^{\eta _1 , \eta _2, \eta _3} _{\eta _0}$, with coefficients which are
$\Sigma$-dependent  2-component spinors $\zeta _{\eta _1, \eta _2, \eta _3}$,
\bea
\label{tendimspin}
\ep = \sum _{\eta _1, \eta _2, \eta _3} \chi ^{\eta _1, \eta _2, \eta _3} \otimes
\zeta _{\eta _1, \eta _2, \eta _3}
\eea
The 10-dimensional chirality condition $\G^{11} \ep = - \ep$ reduces to
\bea
\label{chiral1}
\g_{(4)} \zeta _{- \eta _1, - \eta _2, - \eta _3} =  - \zeta _{\eta_1, \eta _2, \eta _3}
\eea
The Killing spinor equations are invariant under charge conjugation $\chi \to \chi ^c$,
with
\bea
\left ( \chi ^c \right ) ^{\eta _1, \eta _2 , \eta _3}  = B_{(1)} \otimes B_{(2)} \otimes B_{(3)}
\left ( \chi ^{\eta _1 , \eta _2, \eta _3} \right )^*
\eea
Since $ (B_{(1)} \otimes B_{(2)} \otimes B_{(3)})^* = B_{(1)} \otimes B_{(2)} \otimes B_{(3)}$,
and $(B_{(1)} \otimes B_{(2)} \otimes B_{(3)})^2 = I_{16}$, we
may impose, without loss of generality, the reality condition  $\chi ^c = \pm \chi$
on the basis. The sign assignments  are related by (\ref{chiralities}), and
after choosing $\chi^{+++} = + B_{(1)} \otimes B_{(2)} \otimes B_{(3)} (\chi^{+++})^*$
are found to be
\bea
\label{real1}
B_{(1)} \otimes B_{(2)} \otimes B_{(3)}
\left ( \chi ^{\eta _1 , \eta _2, \eta _3} \right )^* = \eta _2
 \chi ^{\eta _1 , \eta _2, \eta _3}
\eea
The $\eta_2$ comes from the fact $B_{(2)}$ anti-commutes with $\g_{(2)}$, while
$B_{(1)}$ and $B_{(3)}$ commute with $\g_{(1)}$ and $\g_{(3)}$ respectively.
Upon imposing the reality condition (\ref{real1}) on the basis of spinors $\chi$,
and the chirality condition (\ref{chiral1}) on $\zeta$, and recalling that
$\chi^{\eta_1 \eta _2 \eta _3}$ has double degeneracy due to the suppressed
quantum number $\eta _0$, we indeed recover 16 complex components for
the spinor $\ep$.

\subsection{Notation}

We introduce a matrix notation in the
8-dimensional space of $\eta$ by,
\bea
\label{taunotation}
\tau ^{(ijk)}  \equiv  \tau^i \otimes \tau ^j \otimes \tau ^k \hskip 1in i,j,k =0,1,2,3
\eea
where $\tau^0=I_2$, and $\tau^i$ with $i=1,2,3$ are the standard Pauli matrices.
Multiplication by $\tau ^{(ijk)}$ is defined as follows,
\bea
(\tau ^{(ijk)} \zeta  )_{\eta _1 , \eta _2 , \eta _3}
& \equiv & \sum _{\eta _1 ', \eta _2 '. \eta _3'}
(\tau ^i)_{\eta _1 \eta _1'} (\tau ^j)_{\eta _2 \eta _2'} (\tau ^k)_{\eta _3 \eta _3'}
\zeta _{\eta _1', \eta _2 '. \eta _3'}
\eea
or more explicitly
\bea
\zeta_{\eta_1, \eta_2, \eta_3} &=& (\zeta)_{\eta_1 ,\eta_2, \eta_3}
\no\\
\zeta_{- \eta_1 ,\eta_2, \eta_3} &=& (\tau^{(100)} \zeta)_{\eta_1, \eta_2, \eta_3}
\no\\
\eta_1 \zeta_{- \eta_1, \eta_2, \eta_3} &=& (+i \, \tau^{(200)} \zeta)_{\eta_1, \eta_2, \eta_3}
\no\\
\eta_1 \zeta_{\eta_1, \eta_2, \eta_3} &=& (\tau^{(300)} \zeta)_{\eta_1 ,\eta_2 ,\eta_3}
\eea
This notation is analogous to the one introduced in \cite{Gomis:2006cu},
and used in \cite{EDJEMG1}.

\newpage

\section{Reduction of the  BPS equations}
\setcounter{equation}{0}
\label{four}

The starting point is the supersymmetry transformation of the gravitino and dilatino
\bea
\delta\lambda
&=& i (\G \cdot P) \cB^{-1} \ep^*
-{i\over 24} (\G \cdot G) \ep
\\
\delta \psi_M
&=& D _M  \ep
+ {i\over 480}(\G \cdot F_{(5)})  \Gamma_M  \ep
-{1\over 96}\left ( \Gamma_M (\G \cdot G)
+ 2 (\G \cdot G) \G^M \right ) \cB^{-1} \ep^*
\no
\eea
In order to preserve 16 supersymmetries, these equation must vanish for sixteen 
independent spinors $\ep$.
For non-constant dilaton ($P \neq 0$), the dilatino equation will reduce the amount of
supersymmetries from 32 to 16, and so the gravitino equation must impose
no additional restrictions.
To proceed, we first reduce these equations using the 
$SO(2,1) \times SO(3) \times SO(5)$ Ansatz.
This will yield algebraic gravitino equations along the directions of the
maximally symmetric spaces, $AdS_2$, $S^2$, and $S^4$.  Using the decomposition
of the ten-dimensional Killing spinor, $(\ref{tendimspin})$, we are able to write the BPS
equations entirely in terms of $\zeta$.  The $\chi^{\eta_1, \eta_2, \eta_3}$
then label the 16 independent supersymmetries.  From the reduced BPS equations,
we obtain a simple set of constraints on bilinears of $\zeta$.  
The net effect of these constraints is to impose a set of projections on $\zeta$.  
This has two benefits, first $\zeta$ is reduced to two complex components
and second using an explicit $SL(2, {\bf R})$ transformation we map the problem to 
one with vanishing axion. That is real dilaton/axion field $P$ and vanishing $Q$.
This is similar to calculations in \cite{EDJEMG1}, \cite{D'Hoker:2006uu}, where the 
problem was also reduced
to one with vanishing axion by an $SL(2, {\bf R})$ transformation.

\subsection{The $SO(2,1) \times SO(3) \times SO(5)$ reduction of the BPS equations}

We give the explicit reduction of the dilatino equation.  The gravitino equations can
be reduced using the same method, so we simply quote the final result.

\smallskip

First we reduce the charge conjugate of the supersymmetry transformation spinor,
\bea
\cB^{-1} \ep^* \sum _{\eta _1, \eta _2, \eta _3} \chi^{\eta_1, \eta_2,\eta_3} \otimes (- i \, \eta_2) 
B_{(4)} \zeta^*_{\eta_1, - \eta_2, \eta_3}
\eea
where we have used the explicit form of $\cB$ given in $(\ref{10dconjugationmarix})$, 
the reality condition on the basis of Killing spinors $\chi^{\eta_1, \eta_2, \eta_3}$ 
given by $(\ref{real1})$, as well as  the chirality condition given by $(\ref{chiralities})$.  
Using this result, the first term in the dilatino equation reduces to
\bea
i (\G \cdot P) \cB^{-1} \ep^* - \sum _{\eta _1, \eta _2, \eta _3}
\chi^{\eta_1, \eta_2, \eta_3} \otimes \bigg( p_a \sigma^a \sigma^2 \tau^{(131)}
\zeta^* \bigg)_{\eta_1, \eta_2, \eta_3}
\eea
We have again used the matrix notation for $\zeta$ introduced in $(\ref{taunotation})$.
The $\tau$-matrices act on the eight-dimensional space spanned by the indices $\{\eta_1, \eta_2, \eta_3\}$,
while the $\sigma$-matrices act on the two-component spinor, $\zeta_{\eta_1, \eta_2, \eta_3}$.
We use a slight abuse of notation, and define $\sigma^{8,9} \equiv \sigma^{1,2}$.
The second term reduces as
\bea
- {i \over 24} (\G \cdot G) \ep
\sum _{\eta _1, \eta _2, \eta _3}  \chi^{\eta_1, \eta_2, \eta_3} \otimes \bigg(
+ {i \over 4} g_a \sigma^a \tau^{(011)} \zeta
- {i \over 4} h_a \sigma^a \tau^{(101)} \zeta
\bigg)_{\eta_1, \eta_2, \eta_3}
\eea
Putting these two terms together yields the dilatino equation.
Since $\chi^{\eta_1, \eta_2, \eta_3}$ spans a basis, the dilatino equation requires 
that the coefficient of each $\chi^{\eta_1, \eta_2, \eta_3}$ vanish separately.  
After dropping the summation and $\chi^{\eta_1, \eta_2, \eta_3}$, and multiplying 
by $\tau^{(131)}$, we have the final form of the reduced dilatino equation,
\bea
(d) & \hskip 0.7in & p_a \gamma^a \sigma^2 \zeta^* + {1 \over 4} g_a \sigma^a \tau^{(120)} \zeta
+ {i \over 4} h_a \sigma^a \tau^{(030)} \zeta = 0
\eea

\smallskip

The gravitino equations are obtained using the same methods.  One additional
step is to replace the covariant
derivative of the spinor along the $AdS_2$, $S^2$, and $S^4$ directions
by the corresponding group action, $SO(2,1)$, $SO(3)$, and $SO(5)$ as defined in
$(\ref{KS})$.  It is important to note that an additional term appears in going
from $\nabla$ to $\hat \nabla$.  This is due to the warp factors appearing in
the ten-dimensional metric.  For example, the covariant derivative along
$AdS_2$ is given by
\bea
\nabla_\mu \ep = \bigg( {1 \over f_1} \hat \nabla_\mu + \half {D_a f_1 \over f_1} \G_\mu \G^a \bigg) \ep
\eea
where $D_a \equiv e_a^M \p_M$ and $M$ is a space-time (Einstein) index.  After a bit of work, we obtain the following
gravitino equations,
\bea
(\mu) &\quad&
0 = - {i \over 2 f_1} \tau^{(211)} \zeta +  {D_a f_1 \over 2 f_1} \sigma^a \zeta
+ \half f_a \sigma^a \tau^{(110)} \zeta
+ { 1 \over 16} \left ( 3 g_a \tau^{(120)} + i h_a  \tau^{(030)} \right ) \sigma^a \sigma^2 \zeta^*
\no \\
(i) &\quad&
0 = + {1 \over 2 f_2} \tau^{(021)} \zeta + {D_a f_2 \over 2 f_2} \sigma^a \zeta
+ \half f_a \sigma^a \tau^{(110)} \zeta
- {1 \over 16} \left ( g_a\tau^{(120)} + 3i h_a  \tau^{(030)} \right ) \sigma^a \sigma^2 \zeta^*
\no \\
(m) &\quad&
0 = + {1 \over 2 f_4} \tau^{(002)} \zeta +  {D_a f_4 \over 2 f_4} \sigma^a \zeta
- \half f_a \sigma^a \tau^{(110)} \zeta
- { 1 \over 16} \left (g_a  \tau^{(120)} - i h_a \tau^{(030)} \right )  \sigma^a \sigma^2 \zeta^*
\no \\
(a) &\quad&
0 = D_a \zeta + {i \over 2} \hat \omega_a \sigma^3 \zeta
- {i \over 2} q_a \zeta + \half f_b \sigma^b \sigma^a \tau^{(110)} \zeta
+ {1 \over 16} \left ( 3 g_a  - g_b \sigma^{ab} \right ) \tau^{(120)}  \sigma^2  \zeta^*
\no\\ && \hskip 2.5in
+ { i \over 16} \left ( -3 h_a + h_b \sigma^{ab} \right ) \tau^{(030)} \sigma^2  \zeta^*
\eea
Here, $\sigma^{ab}$ is  defined by $\sigma^{ab} \equiv \half(\sigma^a \sigma^b - \sigma^b \sigma^a) = i \ep^{ab} \sigma^3$; the derivatives $D_a$ are defined with respect to the 
frame $e^a$, so that $e^a D_a = d$, the total differential on 
$\Sigma$.

\subsection{Symmetries of the reduced BPS equations}

The reduced BPS equations exhibit continuous as well as discrete symmetries,
which will be exploited to further reduce the BPS equations. 

\subsubsection{Continuous symmetries}

The continuous symmetries are as follows.
Local frame rotations of the frame $e^a$ on $\Sigma$ generate a gauge symmetry
$U(1)_c$, whose action on all fields is standard.
The axion/dilaton field $B$ transforms non-linearly under the continuous
$S$-duality group  $SU(1,1)$ of Type IIB supergravity.
As was discussed in section 3.1 of \cite{EDJEMG1}, $B$ takes values in the coset 
$SU(1,1)/U(1)_q$,  and $SU(1,1)$ transformations on the fields are 
accompanied by local $U(1)_q$ gauge transformations, given in section 3.1 of \cite{EDJEMG1},
\bea
U(1)_q & \hskip .6in &
\zeta \to e^{i \theta/2} \zeta 
\no \\
    & \hskip .6in & 
    q_a \to q_a +  D_a \theta     \hskip .7in g_a \to e^{ i \theta} g_a
\no \\
    & \hskip .6in & 
    p_a \to e^{2 i \theta} p_a
     \hskip 0.9in h_a \to e^{ i \theta} h_a
\eea
The real function $\theta$ depends on the  $SU(1,1)$ transformation,
as well as on the field $B$.

\subsubsection{Discrete symmetries}

The reduced BPS equations are invariant under three commuting involutions.
The first two act on $\zeta$ separately from $\zeta ^*$
and leave the fields $f_a, p_a, q_a, g_a, h_a$
unchanged,
\bea
\cI \zeta & = & - \tau ^{(111)} \sigma ^3 \zeta
\no \\
\cJ \zeta & = & \tau ^{(320)} \zeta
\eea
Both $\cI$ and $\cJ$ commute with the symmetries $U(1)_q$ and $U(1)_c$.

\subsubsection{Complex conjugation}

The third involution $\cK$ amounts to complex conjugation. This operation acts 
non-trivially on all complex fields, and its action on $\zeta$ depends on the basis 
of $\tau$-matrices. In a basis in which both $\sigma ^2$  and $\tau^2$ are purely 
imaginary,  the involution $\cK$ has the following form. Taking the complex
conjugates of $p_a, g_a, h_a$, letting $q_a \to - q_a$ and mapping
$\zeta \to  i \tau ^{(012)} \sigma ^2 \zeta ^*$ will leave the BPS equations invariant. 

\smallskip

Complex conjugation, defined this way, however, does not commute with the 
$SU(1,1)$ transformations, since $\zeta$ transforms under $SU(1,1)$ by 
multiplication under a local $U(1)_q$ gauge transformation. Therefore, we 
relax the previous definition of complex conjugation, and allow for complex 
conjugation  modulo a $U(1)_q$ gauge transformation with phase $\theta$,
 \bea
\cK \zeta =  e ^{ i \theta}  \tau ^{(012)} \sigma ^2 \zeta ^*
& \hskip 1in & 
\cK p_a  =  e^{4 i \theta} \, \bar p_a  
\no \\
\cK q_a =  - q_a + 2  D_a \theta \hskip 0.07in
& \hskip 1in & 
\cK g_a = e^{2 i \theta } \, \bar g_a 
\no \\
& \hskip 1in & 
\cK h_a = e^{2 i \theta } \, \bar h_a 
\eea
which continues to be a symmetry of the BPS equations.
The need for such a compenating gauge transformation should be clear from the 
fact that $\zeta $ and $\zeta ^*$ transform with opposite phases under $U(1)_q$.
On the other hand, $\cK$ commutes with the group $U(1)_c$ of frame rotations.

\smallskip

In Type IIB theory only a single chirality is retained, so we have the condition
\bea
\cI \zeta = - \tau ^{(111)} \sigma ^3 \zeta =  \zeta
\eea
This subspace is invariant under the remaining  involutions, since  $\cJ$ and $\cK$ 
commute with $\cI$. 

\newpage

\section{Reality properties of the supersymmetric solutions}
\setcounter{equation}{0}
\label{five}

It is familiar from solving for the Janus solution with 4 supersymmetries in \cite{D'Hoker:2006uu} 
and with 16 supersymmetries in \cite{EDJEMG1} that the BPS equations imply certain reality conditions, which imply that every solution may be mapped into a ``real" solution, for 
which the axion field vanishes. In \cite{EDJEMG1}, these reality conditions were 
derived by first obtaining from the BPS equations  certain bilinear constraints on the 
spinors $\zeta$, and using those to show that only a single eigenspace
of each involution $\cI$, $\cJ$ and $\cK$ should be retained. The reality conditions
for the present problem will be obtained in this manner as well. We repeat
an abbreviated form of the analysis of \cite{EDJEMG1} here, because, even though
the analysis is very similar, its results will be different in subtle but crucial ways.

\subsection{Restriction to a single eigenspace of $\cJ$}

We shall show here that $\zeta$ must obey the projection relation,
\bea
\label{projJ}
\cJ \zeta = \tau ^{(320)} \zeta = \nu \zeta 
\eea
where either the $\nu = +1$ or the $\nu =-1$ eigenspace is retained, but not both.
To derive this result, we start by remarking that the chirality condition implies the 
vanishing of the spinor bilinears $\zeta^\dagger M \sigma^a \zeta = 0$,
for any $\tau$-matrix M which satisfies $\{ \tau^{(111)} \sigma^3, M \sigma^a \}=0$.
Another set of bilinear constraints may be obtained by multiplying
the dilatino equation by $\zeta^\dagger M \sigma^{1,2}$, where
$M$ is a $\tau$-matrix which satisfies
\bea
(T \tau^{(120)})^t = - T \tau^{(120)}
\qquad
(T \tau^{(030)})^t = - T \tau^{(030)}
\eea
The $g_a$ and $h_a$ terms vanish and one is left with
\bea
\zeta^\dagger T \sigma^{1,2} \zeta = 0
\hskip 1.0in
T \in \cT \equiv \{ \tau^{(200)},\tau^{(201)},\tau^{(310)},\tau^{(311)} \}
\eea
We now move onto the gravitino equations.  We first note that we have
$\tau^{(110)} \cT = \cT$.
It then follows that if we multiply the first three gravitino equations by
$\zeta^\dagger T \sigma^{0,3}$, then only the first term in each equation survives and
we obtain
\bea
\zeta^\dagger U \sigma^{0,3} \zeta = 0
\hskip 0.7in
U \in \cU &\equiv& \{
\tau^{(011)}, \tau^{(010)}, \tau^{(101)}, \tau^{(100)}, \tau^{(221)}, \tau^{(220)},
\no\\&&
\tau^{(331)}, \tau^{(330)}, \tau^{(202)}, \tau^{(203)}, \tau^{(312)}, \tau^{(313)}
\}
\eea
These constraints can be solved by first finding a matrix which anti-commutes with 
all of the matrices in $\cU$.  There are two candidates $\tau^{(320)}$ and $\tau^{(231)}$, 
which under multiplication by $\tau^{(111)}$ are equivalent to each other.
The $U$ constraints are automatically satisfied upon imposing the projection condition
(\ref{projJ}); using the methods of Appendix D of \cite{EDJEMG1}, one proves that the  
projection condition (\ref{projJ}) is the general solution to the bilinear constraints.

\subsection{Restriction to a single eigenspace of $\cK$}

We obtain additional constraints by multiplying the $(\mu)$, $(i)$, and $(m)$ gravitino 
equations by $\zeta^\dagger \sigma^p M$, where $p = 0,3$ so that either the $g_a$ 
or $h_a$ term vanishes, but not both.  For the case when $g_a$ survives, 
we choose $M = \tau^{(002)}$, and
for the case when $h_a$ survives, we choose $M = \tau^{(003)}$.  We give explicit
formulas for the first case, and simply quote the results from the second case.  After multiplying
the gravitino equations by $2\zeta^\dagger \tau^{(002)} \sigma^p$,  we have
\bea
\label{Kconst}
0 = - {1 \over  f_1} \zeta^\dagger \tau^{(213)} \sigma^p \zeta
+  {D_a f_1 \over  f_1} \zeta^\dagger \tau^{(002)} \sigma^p \sigma^a \zeta
+  f_a \zeta^\dagger \tau^{(112)} \sigma^p \sigma^a \zeta
+ {3 \over 8} g_a \zeta^\dagger \tau^{(122)} \sigma^p \g^a \sigma^2 \zeta^*
&& \no\\
0 = - {i \over  f_2} \zeta^\dagger \tau^{(023)} \sigma^p \zeta
+  {D_a f_2 \over  f_2} \zeta^\dagger \tau^{(002)} \sigma^p \sigma^a \zeta
+  f_a \zeta^\dagger \tau^{(112)} \sigma^p \sigma^a \zeta
- {1 \over 8} g_a \zeta^\dagger \tau^{(122)} \sigma^p \g^a \sigma^2 \zeta^*
&& \no\\
0 = + {1 \over  f_4} \zeta^\dagger \tau^{(000)} \sigma^p \zeta
+ {D_a f_4 \over  f_4} \zeta^\dagger \tau^{(002)} \sigma^p \sigma^a \zeta
- f_a \zeta^\dagger \tau^{(112)} \sigma^p \sigma^a \zeta
- {1 \over 8}  g_a \zeta^\dagger \tau^{(122)} \sigma^p \g^a \sigma^2 \zeta^*
&&
\eea
For $p=0$, the first three terms are real in the first and third equations and so the 
imaginary part of the fourth term must vanish.  For $p=3$ the first three terms are 
purely imaginary in the second equation, and again the real part of the fourth term 
must vanish.  This gives two bilinear constraints involving $g_a$; listing also the  
corresponding constraints involving $h_a$,
\bea
\label{Kghconstraints}
\Im \bigg( g_a \zeta^\dagger \tau^{(122)} \sigma^a \sigma^2 \zeta^* \bigg) = 0
&\qquad&
\Re \bigg( g_a \zeta^\dagger \tau^{(122)} \sigma^3 \sigma^a \sigma^2 \zeta^* \bigg) = 0
\no\\
\Im \bigg( i h_a \zeta^\dagger \tau^{(033)} \sigma^a \sigma^2 \zeta^* \bigg) = 0
&\qquad&
\Re \bigg( i h_a \zeta^\dagger \tau^{(033)} \sigma^3 \sigma^a \sigma^2 \zeta^* \bigg) = 0
\eea
Taking $p=0$ in the second equation of (\ref{Kconst}), the last three terms are seen 
to be real using the above constraint, while the first term is purely imaginary,
and so the first term must vanish.
For the first and third equations of (\ref{Kconst}), we take $p=3$ and find that the 
first term is real while the last three terms are imaginary and so again, the first term 
must vanish. We quote the corresponding results for the $h_a$ case
\bea
\label{Kconstraints}
\zeta^\dagger \tau^{(022)} \zeta = 0
&\qquad&
\zeta^\dagger \tau^{(023)} \zeta = 0
\no\\
\zeta^\dagger \tau^{(001)} \zeta = 0
&\qquad&
\zeta^\dagger \tau^{(000)} \sigma^3 \zeta = 0
\no\\
\zeta^\dagger \tau^{(212)} \sigma^3 \zeta = 0
&\qquad&
\zeta^\dagger \tau^{(213)} \sigma^3 \zeta = 0
\eea
The  constraints (\ref{Kconstraints}) are solved by imposing a reality condition on $\zeta$,
\bea
\label{complexconj}
\sigma^2 \zeta^*  =  e^{- i \theta } \tau^{(012)} \zeta
\eea
where $\theta$ is an arbitrary phase function on $\Sigma$, which is not fixed by the 
bilinear constraints. This result is readily verified by using (\ref{complexconj}) 
in the form $\zeta ^\dagger = e^{-i \theta} \zeta ^t \sigma ^2 \tau ^{(012)}$
to eliminate $\zeta ^\dagger$ in (\ref{Kconstraints}) and then recognizing that 
the remaining equations are of the form $\zeta ^t M\zeta$ with $M$ anti-symmetric,
and thus vanishes. With the methods used in Appendix D of \cite{EDJEMG1}, 
one demonstrates that (\ref{complexconj}) is in fact the most general solution to 
the bilinear constraints (\ref{Kconstraints}).
The remaining constraints $(\ref{Kghconstraints})$ may be simplified by eliminating 
$\sigma ^2 \zeta ^*$, from (\ref{Kghconstraints}), using (\ref{complexconj}). 
Next, we  use the assumption that $\zeta ^\dagger \tau ^{(130)} \sigma ^a \zeta$
and $\zeta ^\dagger \tau ^{(021)} \sigma ^a \zeta$ are not identically zero (this will
be verified to hold on all the solutions) to obtain,
\bea
\label{realgh}
\Im \left  ( p_a e^{- 2 i \theta} \right ) = \Im \left ( i \,  g_a e^{- i \theta} \right ) \Im \left ( i \, h_a e^{- i \theta} \right ) = 0
\eea
Here we have included also the result of handling the dilatino equation.

\subsection{$SU(1,1)$ map to solutions with vanishing axion}

The arguments that the reality conditions (\ref{realgh}) imply that every solution
to the BPS equations can be mapped to a ``real" solution with vanishing axion
proceed in parallel with the $AdS_4$ case, treated in \cite{EDJEMG1}. We repeat
the keys aspects here for completeness.
The first equation in (\ref{realgh}) implies that the dilaton/axion 1-form $P$
satisfies $P = e^{2 i \theta} \tilde P$, where $\tilde P$ is a real form.
Using the Bianchi identity $ dQ + i P \wedge \bar P=0$, of eq (3.5) of \cite{EDJEMG1}, 
it follows that $dQ=0$, so that $Q$ is pure gauge. Additionally, from the 
$SU(1,1)$ transformation laws (3.13) and (3.14), it follows that the phase $\theta$
is to be interpreted as the accompanying $U(1)_q$ gauge transformation
of an $SU(1,1)$ transformation that maps the solution to the BPS equations
onto a solution for which $\tilde P$ is real, and $Q=0$.
Performing now this $SU(1,1)$ transformation on all fields, allows us 
to set  $e^{- i \theta } =i$, so that the reality conditions (\ref{realgh}) become, 
\bea
\bar p_a = p_a & \hskip 1in & \bar g_a  = g_a \hskip 1in a=8,9
\no \\
q_a=0 & \hskip 1in & \bar h_a  = h_a
\eea
Complex conjugation is now  a symmetry with $\sigma ^2 \zeta ^* =  i  \tau ^{(012)} \zeta$.

\subsection{Reduction to two dimensions}

The projection conditions reduce the number of independent components of $\zeta$
from sixteen complex components to two complex components.  The next step is
to make this reduction explicit in the BPS equations.  The projection conditions are
\bea
\zeta & = & - \tau^{(111)} \sigma^3 \zeta
\no \\
\zeta & = &  \nu \tau^{(320)} \zeta
\no \\
\zeta^* & = & i \sigma^2 \tau^{(012)} \zeta
\eea
In order to implement the first projection, it is convenient to use the 
following rotated basis for the $\tau$-matrices,\footnote{Notice that the transposition and complex conjugation properties of these matrices are identical to those in the 
standard basis.}
\bea
\tau ^1 = \left ( \matrix{1 & 0 \cr 0 & -1 \cr} \right )
\hskip .6in
\tau ^2 = \left ( \matrix{0 & -i \cr i & 0 \cr} \right )
\hskip .6in
\tau ^3 = \left ( \matrix{0 & -1 \cr -1 & 0 \cr} \right )
\eea
The reduction proceeds as follows.  First we denote
the components of $\zeta$ by $\zeta_{\eta_1 \eta_2 \eta_3 \eta_4}$
where $\eta_i = \pm$.  The first projection then fixes the overall 
sign of the $\eta_i$ to be negative.  The next step is to use the second
projection constraint so that the BPS equations contain
$\tau$-matrices whose action preserves
the sign of $\eta_1$.  This guarantees that equations with different
$\eta_1$ indices decouple, and we may fix $\eta_1 = +$.  The equations 
with $\eta_1 = -$ are then automatic.
Using the new basis, the only change
needed is to change the first term in the $(\mu)$ equation to
\bea
+ {i \nu \over 2 f_1} \tau^{(131)} \zeta
\eea
Next we use the reality condition to fix $\eta_4 = -$.  
In order to implement this last projection, we introduce
a chiral basis for $\zeta$.  We now retain the chiral component of each equation,
and use the reality condition which relates $\zeta_\pm$ as
\bea
\zeta_+ = - \tau^{(012)} \zeta_-^*
\eea
to eliminate $\zeta_+$ from the equations.  The equations become
\bea
(d) ~ & \hskip0.3in &
0 = p_z \zeta_- + {1 \over 4} g_z \tau^{(132)} \zeta_-
- {i \over 4} h_z \tau^{(022)} \zeta_-
\no\\
(\mu) ~ &  &
0 = + {i \nu \over 2 f_1} \tau^{(123)} \zeta_-^*
+  {D_z f_1 \over 2 f_1} \zeta_-
+ \half f_z \tau^{(110)} \zeta_-
+ {3 \over 16} g_z \tau^{(132)} \zeta_-
- {i \over 16} h_z \tau^{(022)} \zeta_-
\no\\
(i) ~ &  &
0 = - {1 \over 2 f_2} \tau^{(033)} \zeta_-^*
+  {D_z f_2 \over 2 f_2} \zeta_-
+ \half f_z \tau^{(110)} \zeta_-
- {1 \over 16} g_z \tau^{(132)} \zeta_-
+ {3i \over 16} h_z \tau^{(022)} \zeta_-
\no\\
(m) ~ &  &
0 = - {1 \over 2 f_4} \tau^{(010)} \zeta_-^*
+  {D_z f_4 \over 2 f_4} \zeta_-
- \half f_z \tau^{(110)} \zeta_-
- {1 \over 16}  g_z \tau^{(132)} \zeta_-
- {i \over 16}  h_z \tau^{(022)} \zeta_-
\no\\
(a+) &  &
0 = D_z \zeta_-^*
+ {i \over 2} \hat \omega_z \zeta_-^*
+ f_z \tau^{(110)} \zeta_-^*
- {1 \over 4} g_z \tau^{(132)} \zeta_-^*
- {i \over 4} h_z \tau^{(022)} \zeta_-^*
\no\\
(a-) & &
0 = D_z \zeta_-
- {i \over 2} \hat \omega_z \zeta_-
+ {1 \over 8} g_z \tau^{(132)} \zeta_-
+ {i \over 8} h_a \tau^{(022)} \zeta_-
\eea
These equations are now explicitly decoupled and we can restrict attention
to the two-component spinors $\xi$, whose components we denote
by $\alpha$ and $\beta$,
\bea
\label{xiab}
\zeta _- = \left ( \matrix{ \zeta _{+++-} \cr \zeta _{+---} \cr } \right )
= \left ( \matrix{ \alpha \, \cr  \beta  \, \cr } \right ) & \equiv & \xi
\eea
The key point is that the above $\tau$-matrices transform the above two components
of $\zeta$ denoted by $\xi$ solely into each other.
The action of the $\tau$-matrices on $\xi$ takes the following form
\bea
\tau^{(132)} = - \sigma^2
&\qquad&
\tau^{(123)} = - \sigma^2
\no\\
\tau^{(110)} = + \sigma^3
&\qquad&
\tau^{(033)} = + \sigma^1
\no\\
\tau^{(022)} = - \sigma^1
&\qquad&
\tau^{(010)} = + \sigma^3
\eea
where $\sigma$ are now inthe standard basis of Pauli matrices.
Using this notation, the reduces BPS equations become
\bea
\label{BPS3}
(d) ~ & \hskip0.3in &
0 = p_z \xi - {1 \over 4} g_z \sigma^2 \xi
+ {i \over 4} h_z \sigma^1 \xi
\no\\
(\mu) ~ & &
0 = - {i \nu \over 2 f_1} \sigma^2 \xi^*
+ \half {D_z f_1 \over f_1} \xi
+ \half f_z \sigma^3 \xi
- {3 \over 16} g_z \sigma^2 \xi
+ {i \over 16} h_z \sigma^1 \xi
\no\\
(i) ~ & &
0 = - {1 \over 2 f_2} \sigma^1 \xi^*
+ \half {D_z f_2 \over f_2} \xi
+ \half f_z \sigma^3 \xi
+ {1 \over 16} g_z \sigma^2 \xi
- {3i \over 16} h_z \sigma^1 \xi
\no\\
(m) ~ & &
0 = - {1 \over 2 f_4} \sigma^3 \xi^*
+ \half {D_z f_4 \over f_4} \xi
- \half f_z \sigma^3 \xi
+ {1 \over 16}  g_z \sigma^2 \xi
+ {i \over 16}  h_z \sigma^1 \xi
\no\\
(a+) &  &
0 = D_z \xi^*
+ {i \over 2} \hat \omega_z \xi^*
+ f_z \sigma^3 \xi^*
+ {1 \over 4} g_z \sigma^2 \xi^*
+ {i \over 4} h_z \sigma^1 \xi^*
\no\\
(a-) &  &
0 = D_z \xi
- {i \over 2} \hat \omega_z \xi
- {1 \over 8} g_z \sigma^2 \xi
- {i \over 8} h_a \sigma^1 \xi
\eea
It will be useful to have the transpose of the next to last equation as well,
\bea
0 = D_z \xi^\dagger
+ {i \over 2} \hat \omega_z \xi^\dagger
+ f_z  \xi^\dagger \sigma^3
- {1 \over 4} g_z  \xi^\dagger \sigma^2
+ {i \over 4} h_z  \xi^\dagger \sigma^1
\eea

\subsection{Algebraic relations for the radii $f_1$, $f_2$, and $f_4$}

The reduced gravitino BPS equations contain sectors which are purely
algebraic and may be used to produce algebraic expressions for the radii
$f_1$, $f_2$ and $f_4$ in terms of $\xi$, or equivalently, in terms of $\a$ and $\b$.
The results are as follows,
\bea
\label{radii}
f_1 & = & - \xi ^\dagger \xi  ~~ = - (\bar \a \a + \bar \b \b )
\no \\
f_2 & = & \nu \xi ^\dagger \sigma ^3 \xi = \nu (\bar \a \a - \bar \b \b )
\no \\
f_4 & = & \nu \xi ^\dagger \sigma ^1 \xi = \nu ( \bar \a \b + \bar \b \a )
\eea
The derivation is completely analogous to the derivation given in \cite{EDJEMG1}
for the corresponding radii, and will not be reproduced here. The key in the 
derivation  is to multiply the $(\mu)$, $(i)$, and $(m)$ equations respectively by
$\xi^\dagger$, $\xi^\dagger \sigma^3$, and $\xi^\dagger \sigma^1$, and use
$(a+)$ and $(a-)$ equations to derive the following derivative equations, 
\bea
D_z \left ( \xi^\dagger \sigma^p \xi \right )
= 
- f_z \xi^\dagger \sigma^3 \sigma^p \xi
+ {1 \over 4}  \xi^\dagger \left ( g_z \sigma^2 - i h_z \sigma ^1 \right ) \sigma^p \xi
+ {1 \over 8} \xi^\dagger \sigma^p \left ( g_z \sigma^2 + i h_z \sigma ^1 \right ) \xi
\eea
for $p=0,1,3$. Eliminating $g_z$, $h_z$, and $f_z$ from both sets of equations
yields (\ref{radii}) up to overall multiplicative constants, which are fixed
by using the remaining linearly independent combinations of the $(\mu)$, $(i)$, and $(m)$ equations. 

\subsection{The remaining reduced BPS equations}

Once the relations for the radii (\ref{radii}) have been extracted from the reduced 
BPS equations (\ref{BPS3}), only equations $(d)$, $(a\pm)$ and one particular 
combination of the $(m)$ equations,
\bea
\label{BPS5}
(m) \hskip 1in
0 = - i \nu - i f_z \xi^t \sigma^1 \xi
+ {1 \over 8}  g_z \xi^t \xi
+ {1 \over 8}  h_z \xi^t \sigma^3 \xi
\eea
remain. We may choose conformal coordinates $w, \bar w$ on $\Sigma$, in terms of which
the metric on $\Sigma$ takes the form, $ds^2 _\Sigma = 4 \rho ^2 dw d \bar w$.
The frames, derivatives and connection are then,
\bea
e^z = \rho \, dw  & \hskip 0.5in & 
D_z = \rho ^{-1} \p_w \hskip 1in \hat \o_z = +i \rho ^{-2} \p_w \rho
\no \\
e^{\bar z} = \rho \, d\bar w  & & 
D_{\bar z} = \rho ^{-1} \p_{\bar w} \hskip 1in \hat \o_{\bar z} = -i \rho ^{-2} \p_{\bar w} \rho
\eea
In local complex coordinates, and expressing $\xi$ in terms of $\a$ and $\b$
using (\ref{xiab}), the remaining reduced BPS equations (\ref{BPS5}) take the form,
\bea
(d) & \qquad & 
4 p_z \alpha + i (g_z + h_z) \beta =0
\no \\ && 
4 p_z \beta - i (g_z- h_z) \alpha =0
\no \\
(m) & \qquad &
- i \, \nu - 2 i \alpha \beta f_z 
+ {1 \over 8} (g_z+ h_z) \alpha ^2 + {1 \over 8} (g_z- h_z) \beta ^2 =0
\no \\ 
(a+) & \qquad &
{1 \over \rho} \p_w \bar \alpha   - {1 \over 2 \rho^2} (\p_w \rho)  \bar \alpha 
+ f_z \bar \alpha  - {i \over 4} (g_z- h_z) \bar \beta  =0
\no \\ &&
{1 \over \rho}  \p_w \bar \beta  - {1 \over 2 \rho^2 } (\p_w \rho) \bar \beta 
- f_z \bar \beta  + {i \over 4} (g_z + h_z) \bar \alpha  =0
\no \\ 
(a-) & \qquad &
{1 \over \rho} \p_w \alpha  + {1 \over 2 \rho^2 } (\p_w \rho ) \alpha 
+  {i \over 8} (g_z - h_z) \beta =0
\no \\ &&
{1 \over \rho}  \p_w \beta + {1 \over 2 \rho^2 } (\p_w \rho ) \beta 
  - {i \over 8} (g_z + h_z) \alpha =0
\eea
These equations will be the starting point for the complete solution of the 
reduced BPS equations, to be carried out in the subsequent sections.

\newpage

\section{The BPS equations form an integrable system}
\setcounter{equation}{0}
\label{six}

The exact solution of the $AdS_4$ BPS equations in \cite{EDJEMG1} was obtained
by mapping the BPS equations onto an integrable system which was then 
mapped onto free field equations in turn. The same method also works 
for the problem at hand. The integrable system is very similar, but not
identical, to the one found and used in \cite{EDJEMG1}. The differences between the 
two system will produce key differences between the physical supergravity solutions.
For this reason, and for the sake of completeness, we shall reproduce here  the key 
manipulations required for the integrable system.

\smallskip

First, the dilatino equation $(d)$ may be used to solve for $g_z$ and $h_z$ in terms 
of $\alpha$, $\beta$, $p_z$, 
\bea
g_z & = & 2 i \left ( {\alpha \over \beta} - {\beta \over \alpha} \right ) p_z
\no \\
h_z & = & 2 i \left ( {\alpha \over \beta} + {\beta \over \alpha} \right ) p_z
\eea
in all equations, and the $(m)$ equation may be solved to obtain $f_z$,
\bea
f_z = -{\nu \over 2 \a \b} + {\a^4 - \b^4 \over 4 \a^2 \b^2} p_z 
\eea
The remaining equations may be cast in the following form,
\bea
\label{der}
(a+) & \qquad &
\p_w \ln \left ( { \bar \alpha \over \bar \b} \right )  - {\nu \rho \over \a \b} 
+ {\a^4 - \b^4 \over 2 \a^2 \b^2} \rho  p_z  
   -  \left ( {| \beta |^2 \over |\alpha |^2} - {| \alpha |^2 \over | \beta |^2 } \right ) 
\rho p_z =0
\no \\ &&
\p_w \ln \left ( \bar \a \bar \beta \right )  - \p_w \ln \rho  
 -  \left ( {| \beta |^2 \over |\alpha |^2} + {| \alpha |^2 \over | \beta |^2 } \right ) 
\rho p_z =0
\no \\ 
(a-) & \qquad &
\p_w \alpha + {1 \over 2 \rho} (\p_w \rho ) \alpha 
+  {1 \over 2} { \beta ^2 \over \alpha} \rho p_z =0
\no \\ &&
 \p_w \beta  + {1 \over 2 \rho} (\p_w \rho ) \beta 
+ {1 \over 2}  {\alpha ^2 \over \beta} \rho p_z =0
\eea
Notice that, while there is considerable similarity with the reduced BPS equations
(7.3) and (7.4) of \cite{EDJEMG1}, the details differ. In particular, the coefficients 
of $\rho p_z$ are qualitatively different.

\subsection{Solution of the $(a-)$ system}

Multiplying the first $(a-)$ equation of (\ref{der})  by $2 \rho \alpha$ and the second 
by $2 \rho \beta$, we obtain the equivalent equations, 
\bea
\label{ab1}
\p_w \left ( \rho \alpha^2 \right ) 
+ (\rho p_z) ~ \rho \beta ^2 & = & 0
\no \\
\p_w \left ( \rho \beta ^2 \right ) 
+  (\rho p_z) ~ \rho \alpha ^2 & = & 0
\eea
It follows that $\rho p_z$ is the gradient of a real function, which we shall denote by $\phi$. 
The dilaton field, in standard normalization, is related to $\phi$ by $\phi= \Phi /2$, 
\bea
\rho p_z = \p_w \phi
\eea
Adding and subtracting both equations in (\ref{ab1}), we get 
\bea
\p_w \left ( \ln \left ( \rho (\alpha ^2 + \beta ^2 ) \right ) + \phi \right ) & = & 0
\no \\
\p_w \left ( \ln \left ( \rho (\alpha ^2 - \beta ^2 ) \right ) - \phi \right ) & = & 0
\eea
These equations may be solved in terms of two arbitrary holomorphic  
functions  $\kappa$  and $\lambda$,
\bea
\label{albetsol1}
\rho (\alpha ^2 + \beta ^2) & = & 
\bar \kappa \,  e^{- \bar \lambda  - \phi  }
\no \\
\rho (\alpha ^2 - \beta ^2) & = & 
\bar \kappa \, e^{ + \bar \lambda  +\phi }
\eea
To be more precise, $\lambda$ is a scalar function, but $\kappa$ is a form of
weight $(1,0)$, in a normalization where the frame component $e^z$ is a form 
of weight $(-1,0)$. This gives a complete solution of the $(a-)$ system. 
The product of the two relations in (\ref{albetsol1})  gives,
\bea
\rho ^2 \left (\alpha ^4 - \beta ^4 \right ) = \bar \kappa ^2
\eea
The parametrization in terms of $\kappa$ and $\lambda$ is convenient since 
the following will occur,
\bea
\label{albetsol2}
{\alpha ^2 \over \beta^2} & = & 
{ 1+ e^{2 \phi + 2 \bar \lambda} \over 1- e^{2 \phi + 2 \bar \lambda} }
\no \\
4 \rho ^2 \alpha ^2 \beta ^2 & = &  \bar \kappa ^2 \,  \left (
e^{-2 \phi - 2 \bar \lambda} - e^{2 \phi + 2 \bar \lambda} \right )
\eea
The spinor components $\a$ and $\b$ may be computed as well,
\bea 
\label{ab3}
\a & = & (\bar \kappa /\rho )^{-\half} \ch (\phi + \bar \lambda) ^\half 
\no \\
\b & = & i (\bar \kappa /\rho )^{-\half} \sh (\phi + \bar \lambda) ^\half 
\eea
but this of course has required a choice of signs, which we fix by the above formula.

\subsection{Solution of the $(a+)$ system}

Taking the logarithmic derivatives of the complex conjugates of
equations (\ref{albetsol2}) provides the combinations of derivatives
applied to $\bar \a$ and $\bar \b$ that enter into the $(a+)$ equations,
\bea
\label{logder2}
\p_w \ln \left ( { \bar \alpha  \over \bar \beta  } \right )
& = & 
\half \left ( {\bar \alpha ^2 \over \bar \beta ^2 } - {\bar \beta ^2 \over \bar \alpha ^2 } \right )
\left ( \p_w \phi + \p_w \lambda  \right )
\no \\
\p_w \ln \left ( \rho \bar \alpha \bar \beta  \right )
& = &
\p_w \ln \kappa 
- \half \left ( {\bar \alpha ^2 \over \bar \beta ^2 } + {\bar \beta ^2 \over \bar \alpha ^2 } \right )
\left ( \p_w \phi + \p_w \lambda  \right )
\eea
Eliminating now the logarithmic derivatives between (\ref{logder2}) and the 
$(a+)$ equations, and then eliminating any further algebraic $\a$- and $\b$-dependences
through (\ref{albetsol1}) and (\ref{albetsol2}), we obtain the following first order system,
\bea
\label{inteq1}
\left [ {1 \over \sh (2 \phi + 2 \lambda )} + {1 \over \sh (2 \phi + 2 \bar \lambda )} -
 {2 \, \ch(\lambda - \bar \lambda ) \over | \sh (2 \phi + 2 \lambda ) |}  \right ] \p_w \phi
= {i \sqrt{2}  \nu \rho ^2 \bar \kappa ^{-1} \over \sh (2 \phi + 2 \bar \lambda)^\half } 
- {\p_w \lambda  \over \sh (2 \phi + 2  \lambda )} \qquad
\eea
and 
\bea
\label{inteq2}
 \p_w \ln \rho^2 = \p_w \ln \kappa
+ {\ch (2 \phi + 2 \lambda ) \over \sh (2 \phi + 2 \lambda )}
(\p_w \phi + \p_w \lambda) 
-2 { \ch (2 \phi + \lambda + \bar \lambda ) \over |\sh (2 \phi + 2 \lambda)|}   \p_w \phi 
\eea
This system of first order equations is virtually identical to the one 
encountered in equation (7.12) of \cite{EDJEMG1}; the differences are 
by a replacement of $\sh$ by $\ch$ in the numerator of the third term on the lhs 
of the first equation, and in the numerator of the second term on the rhs
in the second equation, in addition to various signs and factors of $i$.

\subsection{Integrability and the universal dilaton equation}

The above system of first order differential equations is integrable. 
The proof is completely analogous to the proof given for the corresponding
first order system in \cite{EDJEMG1}. Here, we shall limit ourselves to stressing the 
minor differences. A key ingredient in the proof was the derivation of a 
second order partial equation for the dilaton $\phi$ and the holomorphic function $\lambda$ alone. This is also the case here, and the corresponding equations is obtained by  
eliminating $\rho$ between the above first order equations,
\bea
\label{univdilaton}
&&
\p_w  \p_{\bar w} \phi 
- 2 {\sh (4 \phi + 2 \lambda + 2 \bar \lambda) \over | \sh (2 \phi + 2 \lambda)|^2 }
    \p_w \phi \p_{\bar w} \phi 
\no \\ && \hskip .6in 
{\p_{\bar w} \phi \, \p_w  \lambda \over \sh ( \lambda -  \bar \lambda)} \, 
    { \sh (2 \phi + 2 \bar \lambda) \over  \sh (2 \phi + 2 \lambda)}
\left ( - {1 \over \ch (\lambda - \bar \lambda) }  -  
    { \sh (2 \phi + 2  \lambda)^\half  \over  \sh (2 \phi + 2 \bar \lambda)^\half } \right )
    + {\rm c.c.}
\eea
The integrability conditions on $\phi$ and $\ln \rho ^2$ now follows as in \cite{EDJEMG1}.
Note the minor differences between (\ref{univdilaton}) and the corresponding 
equation (7.16)  in \cite{EDJEMG1}.

\newpage

\section{Complete Analytical Solution}
\setcounter{equation}{0}
\label{seven}

In \cite{EDJEMG1}, a judiciously chosen change of variables was discovered, under
which the dilaton equation, analogous to (\ref{univdilaton}), was mapped to an 
equation akin to Liouville and Sine-Gordon field theory, and the first order system, 
analogous to (\ref{inteq1}), is mapped to the corresponding  B\"acklund pair. 
This integrable system, in turn, was mapped onto a set of linear equations, 
to which the full solution may be derived and expressed in terms of two real 
harmonic functions. With minor, but significant, alterations, these changes of 
variables may be adapted to the present system of equations, which may also
be solved completely via these methods.

\smallskip

The change of variables (which coincides with the one carried out  in \cite{EDJEMG1} for the dilaton), 
\bea
\label{theta}
e^{2 i \tet } \equiv {\sh (2 \phi + 2 \lambda) \over \sh (2 \phi + 2 \bar \lambda)}
\eea
maps the dilaton equation (\ref{univdilaton}) to the following equation,
\bea
\label{univ2}
\p_{\bar w} \p_w \tet + {1 \over \sin \mu} 
\left ( e^{-i \tet } \p_{\bar w}  \tet \, \p_w \mu 
    + e^{i \tet } \p_w \tet \, \p_{\bar w} \mu \right )
    - 2 {\cos \mu \over (\sin  \mu)^2}  \p_w \lambda \p_{\bar w} \bar \lambda \sin \tet
    =0
\eea
where we use the notation $\lambda - \bar \lambda = i \mu$, with $\mu$ real.
An alternative form of the equation exploits the special relation that exists
between the second and third terms to recast (\ref{univ2}) as a conservation equation,
\bea
\label{univ3}
\p_{\bar w} \left ( \p_w \tet + 2 i {\p_w \mu \over \sin \mu} e^{- i \tet} \right )
+
\p_w \left ( \p_{\bar w} \tet - 2 i  {\p_{\bar w} \mu \over \sin \mu} e^{+ i \tet} \right )=0
\eea
Equations (\ref{univ2}) and (\ref{univ3}) are clearly again of the Liouville and 
Sine-Gordon type, but differ in subtle ways from the corresponding equation
found in \cite{EDJEMG1}.

\subsection{Changing variables in the first order system}

To simplify the first order system (\ref{inteq1}), we carry out the change of variables 
(\ref{theta}) but, in addition, need to use instead of $\rho$ the new variable $\hat \rho$,
defined by,
\bea
\label{hatrho}
\rho ^8 \equiv {\hat \rho ^8 \over 16} \, \kappa ^4 \bar \kappa ^4
 (\sin 2 \mu )^2\, {\cos \mu - \cos \tet \over (\cos \mu + \cos \tet )^3}
\eea
The first-order system of equations becomes,
\bea
 \p_w \tet & = & - i {\p_w \mu \over \sin \mu} e^{-i\tet } - i \p_w \ln \sin \mu
+ i  \nu \hat \rho ^2 \, \kappa \, e^{ i \tet /2} 
\no \\
  \p_w \ln \hat \rho ^2 & = &  {i \over 2} \p_w \tet 
- {\p_w \mu \over \sin \mu} \, e^{-i \tet } 
\eea
We make a further change of variables, and define
\bea
\label{psi}
\psi \equiv {\sin \mu \over \hat \rho ^2} e^{-i \tet /2}
\eea
It is immediate to recast the first order system in terms of $\Psi$,
\bea
\label{Psieq}
\p_w \psi & = & \nu  \, \kappa \, \sin \mu
\no \\
\p_w \bar \psi & = & \bar \psi  \, {\cos \mu \over \sin \mu} \p_w \mu 
+ \psi \, {\p_w \mu \over \sin \mu} 
\eea
Integrability of this system is now easily checked.

\subsection{Solving the first order system}

We begin by defining the  holomorphic scalar functions $\cA (w)$ and $\cB (w)$ by,
\bea
\label{curlyAB}
\p_w \cA & = & - {\nu \over 2} \kappa \, e^{+ \lambda}
\no \\
\p_w \cB & = &  - {\nu \over 2} \kappa \, e^{ - \lambda}
\eea
up to additive constants. The first equation in (\ref{Psieq}) is readily solved, and we find,
\bea
\psi (w,\bar w) = i e^{- \bar \lambda} (\bar w) \cA(w)
- i e^{\bar \lambda} (\bar w) \cB(w) + \overline{\varphi(w)}
\eea
where $\varphi (w)$ is an as yet to be determined holomorphic function.
Next, we substitute this result into the second equation of (\ref{Psieq}),
and find, 
\bea
&&
2 e^{- \bar \lambda} \bar \cA - 2 e^{\bar \lambda} \bar \cB
-2 e^{- \bar \lambda}  \cA + 2 e^{\bar \lambda}  \cB
\no \\ && \hskip .6in
= i \left ( e^{\lambda - \bar \lambda} - e^{-\lambda + \bar \lambda} \right )
{\p_w \varphi \over \p_w \lambda} 
- i \left ( e^{\lambda - \bar \lambda} + e^{-\lambda + \bar \lambda} \right )
\varphi - 2 i \bar \varphi
\eea
By the same arguments as we used in \cite{EDJEMG1},
the general solution is found to be
\bea
\varphi = -i e^{-\lambda} (\cA - r_1) + i e^\lambda (\cB + r_2)
\eea
where $r_1, r_2 $ are two arbitrary real constants. Assembling now all
contributions to $\psi$, we find,
\bea
\psi = i e^{- \bar \lambda} \left ( \cA + \bar \cA \right )
- i e^{\bar \lambda} \left ( \cB + \bar \cB \right )
\eea
and the constants $r_1, r_2$ have been absorbed into the integration 
constants of $\cA$ and $\cB$ without loss of generality.

\subsection{Solving for the Dilaton}

From the definition of the complex field $\psi$ in (\ref{psi}), it is manifest that
both $\tet$ and $\hat \rho$ may be recovered from $\psi$. In turn, from $\tet$ 
and $\hat \rho$, one derives the $\Sigma$-metric $\rho$ using (\ref{hatrho}), 
and the dilaton $\phi$ using (\ref{theta}), and one derives $\a$ and $\b$ using (\ref{ab3}), 
and ultimately the metric factors $f_1$, $f_2$, and $f_4$ using (\ref{radii}). 
The results are most conveniently expressed in terms of two real harmonic functions
$h_1$ and $h_2$ (on $\Sigma$), which are defined by,
\bea
h_1 & \equiv & \cA + \bar \cA
\no \\
h_2 & \equiv & \cB + \bar \cB
\eea
These steps are all familiar from \cite{EDJEMG1}. In terms of $h_1$ and $h_2$, we have
\bea
\label{firstforms}
\psi & = & i e^{- \bar \lambda} h_1 - i e^{\bar \lambda} h_2
\no \\
\kappa^2 & = & 4 \p_w h_1 \p_w h_2
\no \\
e^{2 \lambda} & = & {\p_w h_1 \over \p_w h_2}
\eea
The function $\tet$ is then given by
\bea
\label{etheta}
e^{i \vartheta} = - { e^{-\lambda} h_1 - e^\lambda h_2
\over  e^{-\bar \lambda} h_1 - e^{\bar \lambda} h_2}
\eea
The dilaton may is computed using  (\ref{theta}), and the third formula 
in (\ref{firstforms}), and we find, 
\bea
\label{dil}
e^{4 \phi} = - \, 
{2 h_1 h_2 |\p_w h_2|^2 - h_2 ^2 (\p_w h_1 \p_{\bar w} h_2 +
\p_w h_2 \p_{\bar w} h_1) 
\over 
2 h_1 h_2 |\p_w h_1|^2 - h_1 ^2 (\p_w h_1 \p_{\bar w} h_2 +
\p_w h_2 \p_{\bar w} h_1)}
\eea
Notice the overall sign difference with the $AdS_4$ dilaton solution of \cite{EDJEMG1}. 
In view of the positivity requirements on $e^{4 \phi}$, this sign difference will have 
drastic effects on the  singularity behavior of the solutions.

\smallskip

A number of combinations of the harmonic functions will be pervasive,
and we shall give them shorthand notations, 
\bea
\label{notation}
V & = & \p _u h_1 \p _{\bar u} h_2 - \p _{\bar u} h_1 \p _u h_2
\no \\
W & = & \p _u h_1 \p _{\bar u} h_2 + \p _{\bar u} h_1 \p _u h_2
\no \\
N_1 & = & 2 h_1 h_2 |\p_u h_1|^2 - h_1 ^2 W
\no \\
N_2 & = & 2 h_1 h_2 |\p_u h_2|^2 - h_2 ^2 W
\eea
The dilaton and metric formulas will take simpler forms, and the regularity conditions on the solutions will be naturally expressed in terms $V,W,N_1$, and $N_2$.

\subsection{Solving for the $\Sigma$-metric}

Combining the definition of $\psi$ in (\ref{psi}), the solution for $\psi$ in (\ref{firstforms}),
the formula for $\tet$ in (\ref{etheta}), and the relation  between $\rho $
and $\hat \rho$ in (\ref{hatrho}), and expressing the result directly in terms of the notation 
(\ref{notation}), we find, 
\bea
\label{rho}
\rho ^8 = - { W^2 N_1N_2 \over h_1^4 h_2^4}
\eea
This equation may be further simplified by including a factor of the dilaton.  Multiplying by 
$e^{\pm 2 \phi}$ we obtain a perfect square on the right hand side.  Taking the square root, we have
\bea
e^{+2 \phi} \rho^4 =  {|W N_2| \over h_1^2 h_2^2}
\hskip 1in 
e^{-2 \phi} \rho^4 =  {|W N_1| \over h_1^2 h_2^2}
\eea
Notice that positivity of $e^{4\phi}$ and of $\rho ^8$ require the same condition
that $N_1N_2<0$.

\subsection{Solving for the Radii}

The radii of $AdS_2$, $S^2$ and $S^4$ are given respectively by (\ref{radii}), namely
\bea
\label{metfacts}
f_1 &=& - (\alpha^* \alpha + \beta^* \beta) \no\\
f_2 &=& \nu (\alpha^* \alpha - \beta^* \beta) \no\\
f_4 &=& \nu (\alpha^* \beta + \beta^* \alpha) 
\eea
It is important to note that the metric factors are explicitly real.  Furthermore, 
$f_1$ is never zero unless $\alpha$ and $\beta$ both vanish; in this case all of 
the metric factors would be zero. The solution $(\ref{ab3})$ for $\alpha$ and 
$\beta$ in terms of $\lambda$ and $\kappa$, gives us formulas for $\alpha$ and 
$\beta$, and thus for $f_1$, $f_2$ and $f_3$.  It is somewhat more convenient
to work with products of metric factors multiplied by the $\Sigma$-metric factor $\rho$,  
\bea
\rho^2  f_1 f_2 &=& - 2 \nu W
\\
\rho^2 f_1 f_4 &=& - 2 \nu \bigg( 
(\p_w h_2)^2 ( (\p_{\bar w} h_2)^2 e^{-4 \phi} - (\p_{\bar w} h_1)^2 ) \bigg)^{\half}
+ {\rm c.c.}
\no\\
\rho^2  f_2 f_4 &=&
2 \bigg( (\p_w h_1)^2 ( (\p_{\bar w} h_2)^2 - e^{4 \phi} (\p_{\bar w} h_1)^2 ) \bigg)^\half
+ {\rm c.c.} 
\eea
Using the dilaton solution (\ref{dil}), the terms under the square roots simplify
because some of their factors are now perfect squares. After some simplifications, 
one obtains,
\bea
\label{signchoice1}
\rho^2 f_1 f_4 &=& - 2 \nu \bigg(   {W  \over N_2 } \bigg)^\half
\bigg( \p_{w} h_2 (h_1 \p_{\bar w} h_2 - h_2 \p_{\bar w} h_1) +s\, 
\p_{\bar w} h_2 (h_1 \p_{w} h_2 - h_2 \p_{w} h_1) \bigg)
\no \\
\rho^2 f_2 f_4 &=&
2  \bigg( {- W    \over N_1}  \bigg)^\half
\bigg( \p_{w} h_1 (h_1 \p_{\bar w} h_2 - h_2 \p_{\bar w} h_1) 
+ s \, \p_{\bar w} h_1 (h_1 \p_{w} h_2 - h_2 \p_{w} h_1) \bigg)
\eea
Here, we have introduced the common sign factor $s=\pm 1$, which arises from 
taking the square roots. 
Reality of $\rho^2 f_1 f_4$ and $\rho^2 f_2 f_4$ forces a correlation between 
the sign $s$ in the above parentheses and the signs of $W, N_1, N_2$.
Recall that to have a well-defined dilaton, we need $N_1$ and $N_2$ to be of 
opposite signs. If $WN_1<0$, then the square roots in both expressions are
taken of positive quantities, which requires the $s=+1$; while if $WN_1>0$,
the square root is taken over a negative quantity, which requires $s=-1$.

\smallskip

With definite sign choices, the formulas simplify considerably, and for the sake of further simplification, we shall present here their squares,  
\bea
N_1 W <0 & \hskip 0.5in & 
\rho^4  f_1^2 f_2^2 = + 4 W^2 
\no \\ &&
\rho^4 f_1^2 f_4^2 = + 4 N_2 W  h_2^{-2}
\no\\ &&
\rho^4 f_2^2 f_4^2 = - 4  N_1 W  h_1^{-2}
\eea
and for the opposite signs,
\bea
N_1 W > 0 & \hskip 0.5in &  \rho^4  f_1^2 f_2^2 = + 4 W^2 
\no \\ &&
\rho^4 f_1^2 f_4^2 =  + 4 h_2 ^2   W  V^2  N_2^{-1}
\no\\ &&
\rho^4 f_2 ^2 f_4^2 = - 4 h_1^2   W V ^2  N_1^{-1}
\eea
From these combinations, we extract the radii themselves,
as well as the products of pairs of radii, which will be very useful later on,
\bea
\label{radii-}
\! \! N_1 W <0 & \hskip 0.6in & 
f_1 ^4 = - 4 e^{+2 \phi} h_1 ^4 \, { W \over N_1}
\hskip 0.9in 
f_2 ^2 f_4^2 = 4 e^{-2 \phi} h_2 ^2
\no \\ &&
f_2 ^4  = + 4 e^{-2 \phi} h_2^4 \, { W \over N_2} 
\hskip 0.9in 
f_1 ^2 f_4^2 = 4 e^{+2 \phi} h_1 ^2 
\no\\ && 
f_4^4 = + 4 e^{-2 \phi}  \, {N_2 \over W}
\hskip 1in 
f_1 ^2 f_2 ^2 
= 4 e^{+2 \phi} {h_1 ^2 h_2^2 W \over N_2} \quad
\eea
and for the opposite signs,
\bea
\label{radii+}
\! \! N_1 W > 0 & \hskip 0.6in & 
f_1 ^4 = - 4 e^{-2 \phi} h_2 ^4 \, { W \over N_2}
\hskip 1in 
f_2^2 f_4^2 = - 4 e^{-2 \phi} h_1^4 h_2^2 {V^2 \over N_1^2}
\no \\ &&
f_2 ^4  = + 4 e^{+2 \phi} h_1^4 \, { W \over N_1} 
\hskip 1in 
f_1^2 f_4^2 = - 4 e^{+2 \phi} h_1 ^2 h_2 ^4 {V^2 \over N_2^2}
\no\\ && 
f_4^4 = + 4 e^{+2 \phi} h_1^4 h_2 ^4 \, {V^4 \over N_1 N_2^2 W}
\hskip 0.5in f_1 ^2 f_2 ^2 
= + 4 e^{-2 \phi} {h_1 ^2 h_2^2 W \over N_1}
\eea
Recall that, since $V$ is purely imaginary, we have $V^2 <0$.

\subsection{The three form fluxes and the Bianchi identities}

It will be helpful to have explicit expressions for the complex gauge potential $B_{(2)}$,
and the associated complex field strength,
\bea\label{AST1}
F_{(3)} = d B_{(2)} = H_{(3)} + i C_{(3)}
\eea 
Its real part $H_{(3)}$ is the NSNS 3-form field strength, while its imaginary part 
$C_{(3)}$ is the  RR 3-form field strength. The Bianchi identities on the 3-forms 
are just the statement that $F_{(3)}$ is closed. These forms will be needed 
to evaluate the corresponding charges. In terms of the 
$AdS_2 \times S^2 \times S^4\times \Sigma$ Ansatz, and conformal gauge on $\Sigma $,
the forms reduce as follows,
\bea
H_{(3)} &=& \hat e^{01} \wedge e^{+ \phi} (g_z \rho dw + g_{\bar z} \rho d \bar w)
\no\\
C_{(3)} &=& \hat e^{23} \wedge e^{- \phi} (h_z \rho dw + h_{\bar z} \rho d \bar w)\label{AST2}
\eea
The forms $\hat e^{01}$ and $\hat e^{23}$ are the unit volume forms respectively
on $AdS_2$ and on $S^2$ and are automatically closed.  The requirement that 
$F_{(3)}$ is closed then implies the local existence of real functions $b_1$ and $b_2$, 
such that
\bea
B_{(2)} & = & b_1 \hat e^{01} + i b_2 \hat e^{23}
\no \\
d b_1 &=& e^{+ \phi} (g_z \rho dw + g_{\bar z} \rho d \bar w)
\no\\
d b_2 &=& e^{- \phi} (h_z \rho dw + h_{\bar z} \rho d \bar w)\label{AST3}
\eea
The calculation of $b_{1,2}$ proceeds in analogy with the corresponding
calculation in \cite{EDJEMG1}. The starting point is obtained by recasting the weight $(1,0)$ part
of the 1-forms $db_{1,2}$ in terms of the dilaton and $\a$ and $\b$,
\bea
e^{+ \phi} f_1^2 \rho g_z &=& 
2 i e^{+ \phi} (\alpha^2 - \beta^2) \bar \alpha \bar \beta 
\bigg( {\alpha \bar \alpha \over \beta \bar \beta} 
+ {\beta \bar \beta \over \alpha \bar \alpha} + 2 \bigg) \p_w \phi 
\no\\
e^{- \phi} f_2^2 \rho h_z &=& 2 i e^{- \phi} (\alpha^2 + \beta^2) \bar \alpha \bar \beta \bigg( {\alpha \bar \alpha \over \beta \bar \beta} + {\beta \bar \beta \over \alpha \bar \alpha} - 2 \bigg) \p_w \phi 
\eea
and expressing these forms as  total derivatives.  The second $(a+)$ equation 
in (\ref{der}) is used to express the combination of the first two terms 
in the large parenthesis on the rhs in terms of a total derivative. Using 
also the complex conjugates of the solutions in (\ref{albetsol2}) to the $(a-)$ equations,
the entire combinations may be expressed as derivatives,
\bea
e^{+ \phi} f_1^2 \rho g_z &=& \p_w \left ( 
2 i \rho ^{-1} \bar \alpha \bar \beta \, \bar \kappa \, e^{+ 2 \phi + \bar \lambda}   \right )
\no\\
e^{- \phi} f_2^2 \rho h_z &=& \p_w \left ( 
2 i \rho ^{-1} \bar \alpha \bar \beta \, \bar \kappa \, e^{- 2 \phi - \bar \lambda}   \right )
\eea
We may now extract $b_{1,2}$ up to some anti-holomorphic functions as
\bea
b_1 &=& 2 i \rho ^{-1} \bar \alpha \bar \beta \, \bar \kappa \, e^{+ 2 \phi + \bar \lambda}  
+ \eta _1(\bar w)
\no\\
b_2 &=& 2 i \rho ^{-1} \bar \alpha \bar \beta \, \bar \kappa \, e^{- 2 \phi  - \bar \lambda}  
+ \eta_2(\bar w)
\eea
Since $b_{1,2}$ must be real, the following combinations must be harmonic,
\bea
2i \rho ^{-1} \bar \a \bar \b  \, \bar \kappa \, e^{\pm (2 \phi + \bar \lambda)}
+ 
2i \rho ^{-1}  \a  \b  \, \bar \kappa \, e^{\pm (2 \phi +  \lambda)}
\eea
and this may indeed be checked explicitly to be the case, in parallel with
\cite{EDJEMG1}. The remaining calculation may be carried out along the lines
of \cite{EDJEMG1} as well. 
We introduce the duals of $h_1$ and $h_2$ denoted by $\tilde h_1$ and $\tilde h_2$, so that
\bea
b_1 = - 2 i {h_1^2 h_2 V \over  N_1} - 2 \tilde h_2
& \hskip 1in & 
\tilde h_1 \equiv i ({\cal A} - \bar {\cal A})
\no\\
b_2 = - 2 i {h_1 h_2^2 V \over  N_2} + 2 \tilde h_1
&& 
\tilde h_2 \equiv i ({\cal B} - \bar {\cal B})
\eea
Note that the overall sign of the 2-form $B_{(2)}$ and its 3-form field
strength $F_{(3)}$ depend on the sign choice made for $\a$ and $\b$
in taking the square roots is (\ref{ab3}).

\subsection{Transformation rules \label{sectiontransformationrules}}

There are a number of simple transformations on the harmonic functions
$h_1$ and $h_2$ which result in simple transformation laws for the dilaton,
metric factors, and 3-form field $g_z, h_z$.
\begin{enumerate}
\item Common scaling $h_{1,2} \rightarrow c^2 h_{1,2}$; leaves the dilaton $\phi$ 
invariant and
\bea
\rho \rightarrow c \, \rho
&&
f_1 \rightarrow c \, f_1
\qquad ~~
f_2 \rightarrow c \, f_2
\qquad ~~~~
f_3 \rightarrow c \, f_3
\no\\
&&
g_z \rightarrow c^{-1}  \, g_z
\qquad
h_z \rightarrow c^{-1} \, h_z
\qquad
f_z \rightarrow c^{-1} \, f_z
\eea
\item Inverse scaling $h_1 \to e^{- \phi_0} h_1$ and $h_2 \to e^{\phi_0} h_2$;
shifts the dilaton, $\phi \to \phi + \phi _0$ and leaves all other fields invariant.
\item Sign reversal $h_1 \to -h_1$, $h_2 \to h_2$, $\nu \to - \nu$; leaves the dilaton and $\rho$
invariant, and 
\bea
\label{h1h2swap}
&&
W \rightarrow - W \qquad
N_{1,2} \rightarrow -N_{1,2}
\no\\&&
f_1 \rightarrow f_1
\qquad ~
f_2 \rightarrow - f_2
\qquad
f_4 \rightarrow f_4
\no\\&&
g_z \rightarrow - g_z
\qquad
h_z \rightarrow h_z
\qquad
f_z \rightarrow - f_z
\eea
\item Interchange $ h_1 \leftrightarrow h_2$; 
\bea
&&
W \rightarrow W
\qquad
N_{1,2} \rightarrow N_{2,1}
\qquad
 \phi \rightarrow - \phi
\no\\&&
f_1 \rightarrow f_1
\qquad
f_2 \rightarrow f_2
\qquad
f_4 \rightarrow f_4^\prime
\no\\ &&
g_z \rightarrow i h_z
\qquad
h_z \rightarrow i g_z
\qquad
f_z \rightarrow f_z^\prime
\eea
The main effect of this transformation is to swap the $S^2$ with $AdS_2$, which 
effectively swaps electric charge with magnetic charge.  
The prime indicates that the fields do not transform simply.  For example for the 
metric factor $f_4$, the exchange of $h_1$ and $h_2$ changes the choice of sign 
for the square root in $(\ref{signchoice1})$. 
\item Lastly, we give the action of the ``S-duality" transformation coming from the 
$SU(1,1)$ symmetry of Type IIB supergravity.  It's action is given by setting 
$\theta = {\pi \over 4}$ in the $U(1)_q$ transformation,
\bea
V: \qquad
e^{+ \phi} \rightarrow e^{- \phi}
\qquad
g_z \rightarrow i g_z
\qquad
h_z \rightarrow i h_z
\eea
The effect here, in contrast to transformation of item 4, is to exchange the NSNS and RR fields.
\end{enumerate}

\newpage

\section{Constant Dilaton Solutions and $AdS_5 \times S^5$}\label{eight}
\setcounter{equation}{0}

Constancy of the dilaton, $\phi = \phi_0$, implies that the solution is $AdS_5 \times S^5$
with $SO(2,4) \times SO(6)$-invariant metric. To show this, it is most convenient 
to analyze the first order equation (\ref{inteq1}), under the assumption $\p_w \phi_0=0$.
This readily gives an expression for the $\Sigma$-metric $\rho$, in terms of the 
constant $\phi_0$ and the holomorphic functions $\kappa$ and $\lambda$,
\bea
\sqrt{2} \nu \rho ^2 = -i \bar \kappa \, \p _w \lambda {\sh (2 \phi_0 + 2 \bar \lambda)^\half 
\over \sh (2 \phi_0 + 2  \lambda)}
\eea
The second first-order equation (\ref{inteq2}) is then automatically solved.
Reality of $\nu \rho^2$ implies the following relation between $\kappa$, $\lambda$
and the constant $\phi_0$,
\bea
-i {\p_w \lambda \over \kappa \, \sh (2 \phi_0 + 2 \lambda)^{3 \over 2}} 
= i {\p_{\bar w} \bar \lambda \over \bar \kappa \, \sh (2 \phi_0 + 2 \bar \lambda)^{3 \over 2}} 
\eea
Since the left hand side is holomorphic, and the right hand side is anti-holomorphic, 
both must be a constant and real. For later convenience, we shall denote this constant by 
$1 /( 2 \sqrt{2} \nu c^2 )$, where $c$ is real.
This equation gives $\kappa$ as a function of $\lambda$ and the constant $c$.
It also gives a formula for $\rho ^2$, and this the $\Sigma$-metric, in terms of $\lambda$ only,
\bea
ds_\Sigma ^2 = \rho ^2 |dw|^2 = { 2 c^2 |d \lambda  |^2 \over |\sh (2 \phi_0 + 2 \lambda)|^2}
\eea
where we have used the fact that $dw \p_w \lambda = d\lambda$, since $\lambda$
is holomorphic. In view of the positivity of $\rho^2$, we must choose $c$ real.
It is natural to change conformal coordinates on $\Sigma$ and use directly
the coordinate in which $ds_\Sigma ^2$ is flat Euclidean,
\bea
ds_\Sigma ^2 = 2 c^2 |dz|^2 
\hskip 1in 
z = - {i \pi \over 2} + \half \ln \left ( {\sh (\phi_0 + \lambda) \over \ch (\phi_0 + \lambda)} \right )
\eea
The shift of $z$ by $- i \pi/2$ is made for later convenience.
Using  (\ref{curlyAB}), and the above expressions for $\kappa$ and 
$\lambda$, now all expressed as functions of the new conformal coordinate $z$,
we readily obtain the  combinations $\p_z \cA$ and $\p_z \cB$, and hence the 
harmonic functions $h_1$ and $h_2$, 
\bea
\label{Ads5zharmonicfunctions}
\p_z \cA = 2 c^2 e^{-\phi_0} \, \sh (z)
& \hskip 1in & 
h_1 =  c^2 e^{-\phi_0} \bigg ( e^z + e^{-z} + e^{\bar z} + e^{- \bar z} \bigg )
\no \\
\p_z \cB = 2 c^2 e^{+\phi_0} \, \ch (z)
& \hskip 1in & 
h_2 =  c^2 e^{+\phi_0} \bigg ( e^z - e^{-z} + e^{\bar z} - e^{- \bar z} \bigg )
\eea
In the sequel of this section, we shall show that this geometry indeed uniquely
corresponds to the $AdS_5 \times S^5$ solution.

\subsection{The $AdS_5 \times S^5$ solution}

At this point, it is useful to express the real harmonic functions $h_1$ and $h_2$
in terms of the real coordinates $x,y$ defined by $z=x+iy$. From (\ref{Ads5zharmonicfunctions}),
we then have
\bea
h_1 &=& 4 c^2  e^{- \phi_0} \, (\ch x) \cos y 
\no\\
h_2 &=&  4 c^2 e^{+\phi_0} \, (\sh x) \cos y 
\eea
We first compute the composite functions defined in (\ref{notation}),
\bea
W &=&  + 4 c^4 \, \sh 2x
\no\\
V &=& - 4 i \, c^4 \sin 2y
\no\\
N_1 &=&  - 64 c^8 e^{- 2 \phi_0} \, (\cos y)^4 \, \sh 2x
\no\\
N_2 &=&  + 64 c^8 e^{+ 2 \phi_0} \, (\cos y )^4 \, \sh 2x
\eea
The dilaton and $\rho$ metric factor are given by
\bea
e^{4 \phi}  = e^{4 \phi_0}
\hskip 1.0in
\rho^4  = 4 c^4
\eea
To evaluate the remaining metric factors, we note that $W$ and $N_1$ have 
opposite signs, and so we must use the first set of metric equations, $(\ref{radii-})$. 
The metric factors are readily computed and found to be given by the customary 
expressions in a coordinate system where the $\Sigma$-metric $\rho$ is constant, 
\bea
\label{ads5s5metricfactors}
f_1^2 &=&  8 c^2 (\ch x)^2
\no\\
f_2^2 &=&  8 c^2 (\sh x)^2
\no\\
f_4^2 &=&  8 c^2 (\cos y)^2
\eea
This is $AdS_5 \times S_5$ with radius $R^2 = 8 c^2$.  
The boundary structure of the space is as follows,  
\begin{itemize}
\item The domain is the half-strip, $0 \leq x \leq \infty$ with $-\pi/2  \leq y \leq \pi/2$
(see figure $2$);
\item $AdS_2$ never shrinks to zero;
\item $S^2$ shrinks to zero when $W$ and $h_2$ are zero at $x=0$;
\item $S^4$ shrinks to zero when $V$, $h_1$ and $h_2$ are zero at $y = -\pi/2, \pi/2$;
\item Throughout the inside of the domain, we have $h_1>0$ and $h_2>0$;
\item $(-N_1)$ and $N_2$ are positive, which guarantees a real dilaton.
\end{itemize}

\begin{figure}[tbph]
\begin{center}
\epsfxsize=4.5in
\epsfysize=3.4in
\epsffile{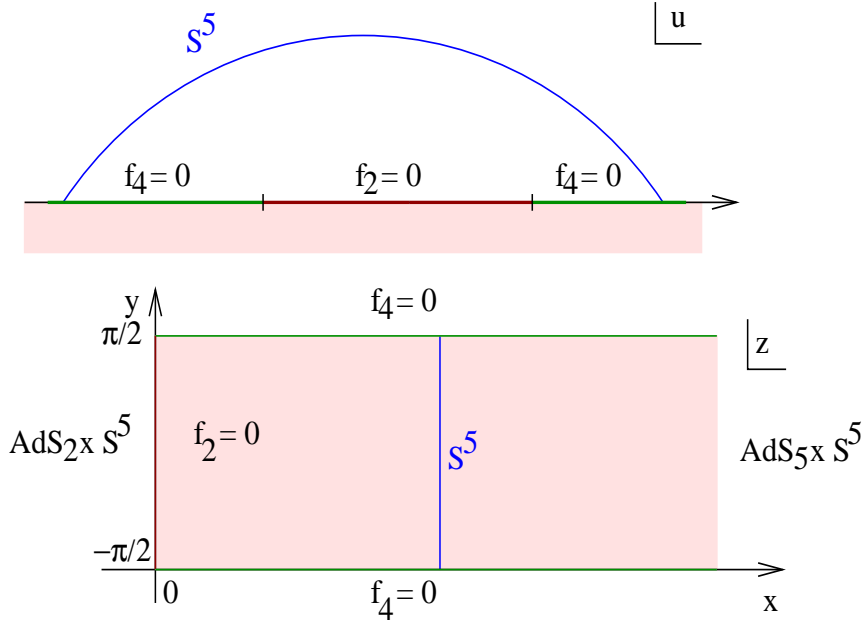}
\label{figure2}
\caption{The $u$-plane (top) and $z$-plane (bottom) for $AdS_5 \times S^5$.}
\end{center}
\end{figure}

\subsection{Mapping the $AdS_5 \times S^5$ solution to the lower half-plane}

To generalize the $AdS_5 \times S^5$ solution, it will be more convenient to
parametrize the solution by the lower half-plane, as in \cite{EDJEMG1}.
The half-strip $0 \leq \Re(w) < \infty$ and $-{\pi \over 2} \leq \Im(w) \leq {\pi \over 2}$ 
is mapped to the lower half-plane $\Im (v) <0$ by the following conformal transformation,
\bea
v = - i \, \sh(z)
\eea
In particular, the imaginary segment from $[- \pi i/ 2, + \pi i / 2]$ in the 
$z$-strip is mapped  to the real segment from $[-1,+1]$ in the $v$-plane.  
The half-line segments  $-i\pi/2 +\bR^+$ and $+ i \pi /2 + \bR^+$ in the 
$z$-strip are mapped respectively to the real axis segments $[-1, -\infty]$ 
and $[+1, +\infty]$ of the $v$-plane.
Finally,  the real segment  $[0, + \infty]$ in the $z$-strip is mapped to the 
negative imaginary axis in the $v$-plane.  

\smallskip

To simplify the expressions, we make use of transformations $1.$ and $2.$ of subsection 
\ref{sectiontransformationrules} to fix $c = 1/\sqrt{2}$ and $\phi_0 = 0$, without loss of generality.
Expressing the harmonic functions (\ref{Ads5zharmonicfunctions}) in terms of the 
$v$-coordinates, we obtain,
\bea
\label{ads5v}
h_1 &=&  \sqrt{1 - v^2}  + {\rm c.c. }
\no\\
h_2 &=& i  (v - \bar v) 
\eea
Note that $h_1$ vanishes along the real axis when $|v| > 1$ and is real when $|v| < 1$.
We shall choose the square root such that $h_1 >0$ for $|v|<1$. The function $h_2$ is 
manifestly positive for $\Im(v) < 0$ and vanishes along the real axis.
The differentials of $h_1 $ and $h_2$ are given by,
\bea
\label{AdS5diff}
\p h_1  &=& - {v dv \over \sqrt{1 - v^2}}
\no\\
\p h_2  &=& i dv
\eea
We see that in the $AdS_5 \times S^5$ solution, the differentials are
both constant multiples of $dv$ as $v \rightarrow \infty$.  This results in a divergence
in the harmonic functions as $v \rightarrow \infty$. 

\smallskip

It is natural to investigate whether the $AdS_5 \times S^5$ solution admits 
simple generalizations of the type achieved by the supersymmetric Janus 
solution in \cite{EDJEMG1}. Upon maintaining the same boundary conditions as 
for $AdS_5\times S^5$, and the 
same branch points and cuts, the only possible generalization is,
\bea
\p h_1  = -  {(v - a)dv \over \sqrt{1 - v^2}}
\eea
while keeping $h_2$ unchanged, for some constant $a$. 
The function $h_1$ is readily obtained by integrating the differential $\p h_1$, 
and we have,,  
\bea
h_1 &=& b +  \sqrt{1 - v^2}  - a \, \arcsin(v) + {\rm c.c. }
\eea
Requiring $h_1$ to vanish at $v = \pm 1$ leads to $a=b = 0$, and the solution
is just $AdS_5 \times S_5$.

\subsection{Preparing for generalization}

In the formulation of the $AdS_5 \times S^5$ differentials given in 
(\ref{AdS5diff}),  there are branch points at $v = \pm 1$, and a double 
pole at $\infty$. Notice that $\infty$ is not a branch point in this formulation.
Actually, there are two different points at $\infty$, one for each Riemann sheet. 
To put this solution in a form which is closer to the Janus solution
and hyperelliptic Ansatz of \cite{EDJEMG2}, we perform a M\"obius 
transformation on the coordinate $v$ to a coordinate $u$, in which 
one of the branch  points, say  $v=-1$, is sent to $u=\infty$, and 
$v=\infty$ is brought back to 2 distinct finite points. 
The M\"obius transformation 
in question is $ u-u_0 = - 1/(v+1)$, and the resulting differentials 
are\footnote{Here, we omit a factor of $1/\sqrt{2}$ in $ \p h_1$,
which may be restored at will using transformations 1. and 2.}
\bea
\label{AdS5}
\p h_1 & = & - i \, {\left ( u - u_0 + 1 \right) du \over (u-u_0)^2 
\sqrt{u-e_1}}
\no \\
\p h_2 & = & i \, {du \over (u-u_0)^2}
\eea
where the branch point is now at $e_1 = u_0 - 1/2$. The lower $v$-half-plane 
is mapped into the lower $u$-half-plane (see figure 2).  The two infinities
in the $v$-half-plane are mapped to the points which we denote by 
$(u_0, +s(u_0))$ and  $(u_0, -s(u_0))$, where $s^2(u) = u - e_1$.

\smallskip

The formulation (\ref{AdS5}) of $AdS_5 \times S^5$ is better suited to 
exhibit the similarities and differences between the $AdS_2$ solution
of this paper and the $AdS_4$ solution of paper \cite{EDJEMG2}. For $AdS_4$,
the differentials $\p h_1$ and $\p h_2$ had double poles at the branch points, 
but nowhere else. For $AdS_2$, the differentials $\p h_1$ and $\p h_2$ are 
regular at the branch points (in fact $\p h_2$ vanishes at both branch points),
but now there are two double poles away from the branch points.
It is the formulation of (\ref{AdS5}) that will be most readily amenable 
to generalization.

\newpage

\section{Conditions for Regular Solutions}\label{nine}
\setcounter{equation}{0}

The general solution to the BPS equations, obtained in section 7 in terms of 
two real harmonic functions $h_1$ and $h_2$ on $\Sigma$, does not always
correspond to acceptable Type IIB geometries.  For example, the harmonic
functions must be chosen so that the right hand side
of the dilaton equation (\ref{dil}) yields a positive or zero value for $e^{\phi}$.
In addition, the metric factors may develop singularities.  For example,
the $AdS_5 \times S^5$ metric factors in (\ref{ads5s5metricfactors}) diverge as $x \rightarrow \infty$.
After mapping $\infty$ to a finite point $u_0$ located on $\p \Sigma$, this divergence shows up
as a double pole in the Abelian differentials given in (\ref{AdS5}).
The gravity dual to a Wilson loop should have only one such asymptotic
$AdS_5 \times S^5$ region.  Correspondingly, we will allow only
one such $AdS_5 \times S^5$ singularity located on $\p \Sigma$.

\smallskip

We restrict attention to solutions where $N_1$, $N_2$ and $W$ have definite sign
throughout the domain $\Sigma$.  If this were not the case, these quantities would
develop zeros on the inside of $\Sigma$, which would -- generically -- lead to 
singularities in the dilaton, or in the radii $f_1$, $f_2$ or $f_4$.  
Without loss of generality, transformation $3.$ of section 7.7 may be used to choose $W$ positive.  
Reality of the dilaton $\phi$ and positivity of $\rho ^2$ then require that 
$N_1$ and $N_2$ have opposite signs. 
We now make the additional assumption that $W N_1<0$, so that the expressions
for the metric factors in (\ref{radii-}) are valid.  We repeat them here for convinence
\bea
\label{radii-2}
f_1^2 f_4^2 & = & 4 e^{2 \phi} h_1^2
\no \\
f_2^2 f_4^2 & = & 4 e^{-2 \phi} h_2^2
\no \\
f_1^4 & = & -e^{2 \phi} h_1^4 {W \over N_1}
\eea
Using transformation $4.$,
it is possible to map
solutions with $W N_1>0$ to solutions with $W N_1 < 0$, but this map does not
guarantee that regular solutions are mapped to regular solutions.  In 
particular, the problem
of generating regular solutions in the case $W N_1>0$ is an open question.

\subsection{Topology conditions}

Since we are interested in the gravity duals of Wilson loops, we require the
ten-dimensional geometry to have a boundary with the topology of $AdS_5 \times S^5$.
This means that every point in $\p \Sigma$, except for the $AdS_5 \times S^5$ singularity mentioned
in the first paragraph, must correspond to a regular interior point in the ten-dimensional geometry.
This is achieved by requiring either $S^2$ or $S^4$ to shrink to zero size on $\p \Sigma$.
The product $f_2^2 f_4^2$ given in (\ref{radii-2}) vanishes if and
only if $h_2$ vanishes (assuming a regular dilaton).  This means the boundary is 
specified completely by $h_2=0$,   
\bea
\p \Sigma = \left \{ \Im (v)<0 ~ {\rm such ~ that} ~ h_2(v)=0 \right \}
\eea
We make the additional requirement that the product $f_2^2 f_4^2$ vanishes on a single line
so that $\p \Sigma$ has a single component.
It is useful to choose conformal coordinates on the lower half-plane in which 
$h_2 = - 2 \Im (v)$; since $h_2$ is a harmonic function to begin with, this can always be done.
These are the coordinates used in (\ref{ads5v}) for $AdS_5 \times S^5$.
Note that $h_2$ is positive throughout the lower half-plane.
The product $f_2^2 f_4^2$ vanishes automatically on the real axis, which
is now the boundary, $\p \Sigma$. 

\smallskip

Using the restrictions of the first three paragraphs of this section, namely the 
$AdS_5 \times S^5$ singularity restriction from the first paragraph, the sign 
choices of the second paragraph, and the boundary $AdS_5 \times S^5$ 
topological restrictions 
of the third paragraph, we derive a set of restrictions on the harmonic functions.
We first compute a few useful quantities in the 
well-adapted coordinates $v=x+iy$, and enforce the sign choices, 
\bea
\label{WN1N2}
W &=& - \p_y h_1 \geq 0
\no\\
N_1 &=& - h_1 \bigg( y (\p_x h_1)^2 + y (\p_y h_1)^2 + h_1 W \bigg) \leq 0
\no\\
N_2 &=& - 4 y \bigg( h_1 - y W \bigg) \geq 0
\eea
The metric factor $f_1$ never vanishes.  This follows from (\ref{metfacts}),
where it is obvious that $f_1$ may vanish if and only if $\alpha$ and $\beta$ vanish.  
The metric products in (\ref{radii-2}) imply the
harmonic functions $h_1$ and $h_2$ may not contain singularities, except for the $AdS_5 \times S^5$ singularity located on $\p \Sigma$.  Finally, the product for $f_1 f_4$ given in (\ref{radii-2}) implies that $f_4$ vanishes
whenever $h_1$ vanishes, which may only happen on $\p \Sigma$.  Given the sign choice $W>0$ and the explicit
form in (\ref{WN1N2}) it follows that $h_1$ is positive throughout the lower half-plane.

\subsection{Summary of Topology and regularity conditions}

We summarize the restrictions as follows.
\begin{description}
\item[(R1)] The harmonic functions {\sl $h_1$ and $h_2$ are non-singular except for one point on $\p \Sigma$ corresponding to the asymptotic $AdS_5 \times S^5$ region.}
\item[(R2)] The boundary {\sl $\p \Sigma$ is defined by the line $h_2 = 0$.}
\item[(R3)] The function {\sl $h_1$ may vanish only on the segment of $\p \Sigma$ 
where $S^4$ shrinking to zero.  The regions
of $\p \Sigma$ where $h_1 \neq 0$ correspond to $S^2$ shrinking to zero.}
\item[(R4)] The functions {\sl $h_1$ and $h_2$ are positive definite inside $\Sigma$, 
but may vanish on  $\p \Sigma$.}
\end{description}

\smallskip

Given that $h_1$ vanishes only on $\p \Sigma$, it follows from (\ref{WN1N2}) that 
$N_2$ is a sum of two positive quantities and vanishes only on $\p \Sigma$.  
In order for the dilaton to be regular, $N_1$ may vanish only when $N_2$ does, and
so only vanishes on $\p \Sigma$. Finally, in order for $f_1$, given in (\ref{radii-2}), 
to be regular and non-zero in the bulk of $\Sigma$, 
$W$ may vanish only when $N_1$ vanishes, and so only vanishes on $\p \Sigma$ as well.
Given that $h_1$, $h_2$, $W$, $N_1$ and $N_2$ never vanish in the bulk of $\Sigma$,
it can be verified that
the equations for the dilaton in (\ref{dil}),
the metric factor $\rho$ in (\ref{rho}), and the 
metric factors $f_1$, $f_2$, and $f_4$ in (\ref{radii-2})
all give regular results.  The requirements on $W$, $N_1$ and $N_2$ may be condensed to the single additional
requirement
\begin{description}
\item[(R5)] {\sl $(-W N_1) > 0$ throughout $\Sigma$.}
\end{description}
This follows from the fact $W$ or $N_1$ may vanish if and only if their product vanishes.  Secondly,
$N_2$ may vanish only if $W$ vanishes.  

\subsection{Dirichlet or Neumann conditions and regularity on $\p \Sigma$}

Remarkably, the non-linear inequalities of (\ref{WN1N2}) admit a linearization,
similar to the one derived in \cite{EDJEMG2}. There, the boundary condition 
$W=0$ was decomposed into a sequence of alternating Neumann and Dirichlet 
boundary conditions for the harmonic functions $h_1$ and $h_2$. An analogous
mechanism will now be identified for the problem at hand.

\smallskip

To this end, we assume that $h_1$ admits a Taylor expansion in 
powers of $y$ away from the boundary at $y=0$.
We have the following expansion,
\bea
h_1 = a_0(x) + a_1(x) y + a_2(x) y^2 + a_3(x) y^3 + \cO(y^4)
\eea
where $a_0,a_1,a_2,a_3$ are real functions of $x$. Since $h_1$ is harmonic, it satisfies 
$(\p_x^2 + \p_y^2) h_1 = 0$. It follows that $a_0(x)$ completely determines the 
coefficients of the terms with even powers of $y$, while $a_1(x)$ completely 
determines the coefficients of the terms with odd powers of $y$; we have 
$2 a_2(x) = - \p_x^2 a_0(x)$ and $6 a_3(x) = - \p_x^2 a_1(x)$.

\smallskip

Since $N_2$ automatically vanishes at $y=0$, regularity of the dilaton at $y=0$
requires that also $N_1$ vanishes at $y=0$.  
This requirement may be enforced on the Taylor expansion of $N_1$, 
\bea
N_1 = -{1 \over 4} a_0^2 a_1 
- {1 \over 4} \bigg( a_0 a_1^2 + 2 a_0^2 a_2 - a_0 (\p_x a_0)^2 \bigg) y + \cO(y^2)
\eea
We see that we must have either $a_0 = 0$ or $a_1 = 0$. Equivalently, at $y=0$, 
the harmonic function $h_1$ satisfies either $\p_y h_1 = 0$, or $h_1 = 0$, i.e. it 
satisfies either Neumann or {\sl vanishing} Dirichlet boundary conditions on $\p \Sigma$.
This result is analogous to the boundary value problem derived  in the $AdS_4$ case
in \cite{EDJEMG2}.

\smallskip

We now verify both sets of boundary conditions yield regular solutions on $\p \Sigma$.
This may be verified directly from the Taylor expansion. In the case of 
Neumann boundary conditions,  $W$, $N_1$ and $N_2$ have 
the following expansions,
\bea
W &=& 2 a_2 y + \cO(y^3)
\no\\
N_1 &=&  -\half a_0^2 a_2 y  + {1 \over 4} a_0 (\p_x a_0)^2  y + \cO (y^3)
\no\\
N_2 &=&  {1 \over 16} a_0  y  + \cO(y^3)
\eea
In the case of {\sl vanishing} Dirichlet boundary conditions, with $a_0 = 0$, 
$N_1$ and $N_2$ have the following expansions,
\bea
W &=& a_1 + \cO(y^2)
\no\\
N_1 &=& \half a_1^2 a_3 y^4 + {1 \over 4} a_1 (\p_x a_1)^2  y^4 + \cO (y^6)
\no\\
N_2 &=& -  {1 \over 8} a_3 y^4 + \cO (y^6)
\eea
Using either set of expressions, it can be verified that
the equations for the dilaton in (\ref{dil}),
the metric factor $\rho$ in (\ref{rho}), and the 
metric factors $f_1$, $f_2$, and $f_4$ in (\ref{radii-})
all give regular results in the limit $y \rightarrow 0$.

\newpage

\section{Hyperelliptic Solutions}\label{ten}
\setcounter{equation}{0}

In this section, we shall present the general hyperelliptic form of regular 
solutions to the BPS equations. From the point of view of the Riemann surface
$\Sigma$, the $AdS_5 \times S^5$ solution corresponded to $\Sigma$
being a (half-) plane, and the harmonic functions $h_1$ and $h_2$ obeying
alternating Neumann and vanishing Dirichlet boundary conditions. 
The construction to be given in this section will generalize $\Sigma$
to a hyperelliptic Riemann surface of general genus with boundary, where the 
harmonic functions obey generalized alternating Neumann and vanishing Dirichlet
boundary conditions.

\smallskip

The starting point of the construction will be a hyperelliptic Riemann 
surface of arbitrary genus $g$, with real branch points $e_1, e_2 , \cdots , 
e_{2g+1}, e_{2g+2} = \infty$, and the following associated hyperelliptic curve,
\bea
s^2 = (u - e_1) \prod _{i=1} ^g (u- e_{2i}) (u - e_{2i+1})
\eea
For later convenience, we choose the branch points
to be ordered as follow,
\bea
e_{2g+1} < e_{2g} < \cdots < e_2 < e_1
\eea
Physical considerations underly the choice of possible differentials
$\p h_1$ and $\p h_2$. 

\smallskip

First and foremost, the gravity dual to a 
Wilson loop should have only a single asymptotic $AdS_5 \times S^5$ region.
In the formulation (\ref{AdS5}) of $AdS_5 \times S^5$, this corresponds
to the two double poles at $(u_0, \pm s(u_0))$. Thus, we shall require
that the differentials $\p h_1$ and $\p h_2$ in the hyperelliptic generalization
have precisely the same double pole structure as $AdS_5 \times S^5$ had.
We shall without loss of generality keep the ordering $e_1 < u_0$.

\smallskip

Second, the harmonic function $h_2$ may be used for a description 
of the boundary of $\Sigma$. As argued in the previous section, 
throughout the boundary, we have $h_2=0$, and this equation in 
turn describes the boundary completely. Thus, we shall leave $h_2$
unmodified from its $AdS_5 \times S^5$ expression in (\ref{AdS5}).

\smallskip

Third, the differential $\p h_1$ should be generalized in such a way 
that it remain regular, except at the double poles $u_0$. Assembling 
all of the above requirements, we arrive at the following 
differentials,
\bea
\label{hyper}
\p h_1  & = & -i \, {P(u) \, du \over (u-u_0)^2 \, s(u)}
\no\\
\p h_2 & = &  i \, { d u \over (u - u_0)^2}
\eea
Here, $P(u)$ is a polynomial in $u$. Since $h_1$ will obey alternating
Neumann and vanishing Dirichlet boundary conditions on $\p \Sigma$,
represented here by the real line $\Im (u)=0$, the polynomial $P$
must have either only imaginary or real coefficients. Without loss of generality,
we may choose the coefficients to be all real. Since $\p h_1$ should 
be regular at the branch point at $\infty$, the polynomial $P$
must have degree at most $g+1$. In analogy with the $AdS_5 \times S^5$
solution, $\p h_1$ should be non-vanishing at $\infty$, so that $P(u)$
must be exactly of degree $g+1$.

\subsection{Counting parameters}

At this point, it is useful to count the number of parameters of 
the hyperelliptic Ansatz. By the regularity arguments exhibited in 9.1, 
the function $h_1$ is required to vanish on the real axis in the regions 
with Dirichlet boundary conditions. There are exactly $g + 1$ such regions 
which gives $g$ period relations.
The polynomial $P$ has $g + 1$ free parameters, while there are $2g -1$ 
parameters $e_i$ (up to overall $SL(2,\bR)$ conformal rotations of $\Sigma$) 
and the parameter $u_0$, yielding a total of $3 g + 1$ parameters.  
After solving the $g+1$ constraints, this leaves $2 g$ 
free parameters.  In addition, there are 3 parameters for $SU(1,1)$ rotations 
to the general solution with axion, 1 for the overall dilaton shift, and 1 for the overall radius.
The total is $2 g + 5$.  When $g=0$, the dilaton and axion are constant so there 
are a total of  only 3 parameters. This counting agrees with the result of 
subsection 8.2 that $AdS_5 \times S^5$ is the only non-singular solution
with $g =0$.

\subsection{Regularity conditions, $ W>0$}

In this subsection, we shall enforce two of the regularity conditions, 
arrived at in section 9, namely $W>0$ and $h_1, h_2 >0$. From 
(\ref{hyper}), we readily evaluate $W$ to be,
\bea
W = - { 1 \over | u- u_0 |^4} \left ( {P(u) \over s(u)} + {\rm c.c.} \right )
\eea
Examining the behavior of $W$ in the neighborhood of a putative 
complex zero $u_i$ of $P(u)$, with $\Im (u_i)<0$, 
it is clear that $W$ cannot maintain
a constant sign circling around $u_i$. (This argument was made
fully explicit in \cite{EDJEMG2}.) Thus, all $g+1$ zeros of $P(u)$ must be real,
and we shall denote them by $\a _b$ for $b=1,\cdots, g+1$, so that
\bea
P(u) = \prod _{b=1 } ^{g+1} (u - \a _b)
\eea 
Reality of the zeros $\a_b$ does not by itself guarantee positive $W$.
To enforce positivity of $W$ on the real axis, we need the behavior 
of $s(u)$ on the the real axis. Clearly, the phase of $s(u)$ changes
by a factor of $i$ upon traversing any branch point. This behavior is
given by
\bea
\label{sphase}
s(u)/|s(u)| & = & -1  \hskip 0.5in 
u \in \cU_1 \equiv ]e_1, +\infty[ \, \cup \, \bigcup _{j=1}^{n_1}
]e_{4j+1} , e_{4j}]
\no \\
s(u)/|s(u)| & = & +i  \hskip 0.5in 
u \in \cU_2 \equiv \bigcup _{j=0}^{n_2}
]e_{4j+2} , e_{4j+1}[
\no \\
s(u)/|s(u)| & = & +1  \hskip 0.5in 
u \in \cU_3 \equiv  \bigcup _{j=0}^{n_3}
]e_{4j+3} , e_{4j+2}]
\no \\
s(u)/|s(u)| & = & -i  \hskip 0.5in 
u \in \cU_4 \equiv \bigcup _{j=0}^{n_4}
]e_{4j+4} , e_{4j+3}[
\eea
The upper limits of these unions are given by
\bea
\label{nvals}
n_1 & = & [g/2] 
\no \\
n_2 & = & [(2g-1)/4] + n_{-\infty}
\no \\
n_3 & = & [(g-1)/2]
\no \\
n_4 & = & [(2g-3)/2] + 1- n_{-\infty}
\eea
where $n_{-\infty} =1$ when $g$ is even and $n_{-\infty}=0$ when $g$ is odd,
and $e_{2g+2}=-\infty$.
The analysis of $W>0$ on the real axis proceeds as follows. 

\smallskip

When $u \in \cU_2$ or $u \in \cU_4$, $s(u)$ is imaginary while $P (u)$ is real,
so that we have $W=0$ automatically  on those segments.  Using the analysis 
of section 9.2, this is the boundary component on which the sphere $S^2$ 
shrinks to zero; $W>0$  places no restrictions here. 

\smallskip

When $u \in \cU_1$, or $u \in \cU_3$ both $s(u)$ and $P(u)$ are real, 
and $W>0$ imposes the conditions,
\bea
u \in \cU_1 & \hskip 1in & P(u) \geq 0
\no \\
u \in \cU_3 & \hskip 1in & P(u)  \leq 0
\eea
This means that $P(u)$ changes sign across each interval in $\cU_2$
and each interval in $\cU_4$, so that it must have an odd number of
zeros in each of these intervals. The total  number of intervals in 
$\cU_2 \cup \cU_4$ is precisely $g+1$. Since $P(u)$ has at most $g+1$ zeros,
we find  that $P(u)$ has no zeros in $\cU_1 \cup \cU_3$
and has precisely one zero in each interval in $\cU_2 \cup \cU_4$.
Thus, we have a {\sl unique ordering of the branch points and zeros of $P(u)$},
given by, 
\bea
\label{order}
\a _{g+1} < e_{2g+1} <  \cdots <
e_{2b+1} < e_{2b} < \a _b < e_{2b-1} < e_{2b-2} < \cdots <
 e_2 < \a_1 < e_1 < u_0
\eea
for $b=2, \cdots , g$.
This ordering is a necessary condition for $W>0$.
We have no general proof that the ordering conditions are also sufficient,
though they will appear to be in the elliptic case, to be treated in
detail in the subsequent section.

\subsection{Regularity conditions, $h_1, h_2 >0$}

The function $h_2$ is readily obtained by integrating the 
differential $\p h_2$ and enforcing the boundary condition $h_2=0$
at $\p \Sigma$ and we find,
\bea
h_2 (u) = - {i \over u-u_0} +  {i \over \bar u - u_0} = {i (u- \bar u) \over |u-u_0|^2}
\eea
and this is manifestly positive in the lower half $u$ plane. Positivity of 
$h_1$ requires that $\Im (\p h_1) <0$ along the $h_1$ Dirichlet intervals 
$\cU_1$ and $\cU_3$. This condition precisely coincides with the 
condition $W>0$, as is also clear from the form of $W$ given in
(\ref{WN1N2}). 

\subsection{Vanishing Dirichlet conditions}

By construction, the function $h_2$ vanishes all along the boundary $\p \Sigma$.
The only vanishing Dirichlet condition which is not automatic is on $h_1$.
It is enforced by the vanishing of the following line integrals that join consecutive 
Dirichlet segments for $h_1$, 
\bea
 \int _{e_{2j}} ^{e_{2j-1}} \p h_1  =0 
\hskip 1in j= 1, \cdots , g+1
\eea
where $e_{2g+2}=- \infty$. These integrals are automatically real
since by definition $\p h_1$ has an overall factor of $i$ in (\ref{hyper}),
and on the integration intervals, $s(u)$ is purely imaginary.

\subsection{Homology 3-spheres and RR 3-form charges}

The $g$ homology 3-spheres of the genus $g$ solutions are given by
\bea
S_b ^3 = [e_{2b+1}, e_{2b} ] \times _f S^2 \hskip 1in b=1,\cdots , g
\eea
and the associated RR 3-form charge is given by
\bea
\cC_b = \int _{S_b^3} C_{(3)} = 8 \pi \int _{e_{2b+1}} ^{e_{2b}} d \tilde h_1
= 8\pi i \int _{e_{2b+1}} ^{e_{2b}} \p  h_1 + {\rm c.c.}
\eea
This allows us to write down an explicit formula for these charges,
\bea
\cC_b = 16 \pi \int _{e_{2b+1}} ^{e_{2b}}  {P(u) \, du \over (u-u_0)^2 \, s(u)} 
\eea
These integrands are manifestly real and of uniform sign, and thus the charges are real
and non-vanishing.

\newpage

\section{Elliptic Solutions}
\label{eleven}
\setcounter{equation}{0}

In this section, we work out the genus 1 case in complete detail, and 
show that regular solutions do indeed exist. We shall also evaluate 
their 3-form and 5-form charges and identify the corresponding fluxes.

\smallskip

The starting point for the genus 1 construction  is the elliptic curve, 
and its uniformization by the  Weierstrass $\wp$-function,
\bea
s(u) ^2 & = & (u-e_1)(u-e_2)(u-e_3)
\no \\
\wp '(z) ^2 & = & 4 (\wp (z) - e_1) (\wp (z) - e_2) (\wp (z) - e_3)
\eea
Here, we have $u = \wp (z)$ and the sign of $s(u)$ is chosen so that 
$2s(u) =  \wp '(z)$. Without loss of generality,  we may translate $u$ 
and the branch points $e_i$ by an overall constant, and choose $e_1+e_2+e_3=0$. In terms of the half-periods $\omega _1$ (real), 
$\omega _2= \omega _1 + \omega _3$, and $\omega _3$ (purely imaginary), 
we have $e_i = \wp (\omega_i)$ for $i=1,2,3$. 
We shall also make use of the Weierstrass $\zeta (z)$- and 
$\sigma (z)$-functions, which are defined, in terms of $\wp (z)$ by 
\bea
\zeta '(z) & = & - \wp (z)
\no \\
\zeta (z) & = & \sigma '(z) /\sigma (z)
\eea
and the requirement that $\sigma (z) = z + \cO (z^3)$.

\subsection{Elliptic differentials}

For the genus 1 case, $P(u)$ has two real zeros, $\a_1$ and $\a_2$,
subject to the  ordering condition,
\bea
\a_2 < e_3 < e_2 < \a_1 < e_1 < u_0
\eea
To completely integrate the elliptic case, it will turn out to be convenient
to use the following alternative parametrization of $P(u)= (u-\a_1)(u-\a_2)$,
\bea
P(u) & = &  (u-u_0)^2 + B_1 (u-u_0) + B_2
\no \\
B_1 & = & 2 u_0 - \a_1 - \a_2
\no \\
B_2 & = & (u_0-\a_1) (u_0 - \a_2)
\eea
The constants $B_1$ and $B_2$ are real, and in view of the relative ordering 
of the branch points and zeros, we have $B_1>0$ and $B_2>0$.
The differentials $\p h_1$ and $\p h_2$ are then uniquely specified by $\a_1, \a_2$ and $u_0$, and given by,
\bea
\p h_1 & = & -i \, {(u-u_0)^2 + B_1 (u-u_0) + B_2 \over (u-u_0)^2} \, 
{du \over  s(u)}
\no \\
\p h_2 & = & i \, { du \over (u-u_0)^2}
\eea
Given that $u_0$ corresponds to the $AdS_5 \times S^5$ limiting region,
$u_0$ must be real and lie in the interval $[e_1, +\infty]$. We define the 
point $ w_0$ by $u_0 = \wp (w_0)$ and use the fact that $u_0 \in [e_1, +\infty]$
to further locate $w_0 \in [0, \omega _1]$. The differentials may then be 
expressed in terms of the uniformization coordinates $z$,
\bea
\label{ellipticansatz}
\p h_1  & = &  
-2 i \bigg( {B_2 \over (\wp(z)- \wp(w_0))^2} + {B_1 \over \wp(z) - \wp(w_0)} + 1 \bigg) dz
\no\\
\p h_2  & = & 
+ i  {\wp^\prime(z) dz \over (\wp(z) - \wp(w_0))^2}
\eea

\begin{figure}[tbph]
\begin{center}
\epsfxsize=5in
\epsfysize=2in
\epsffile{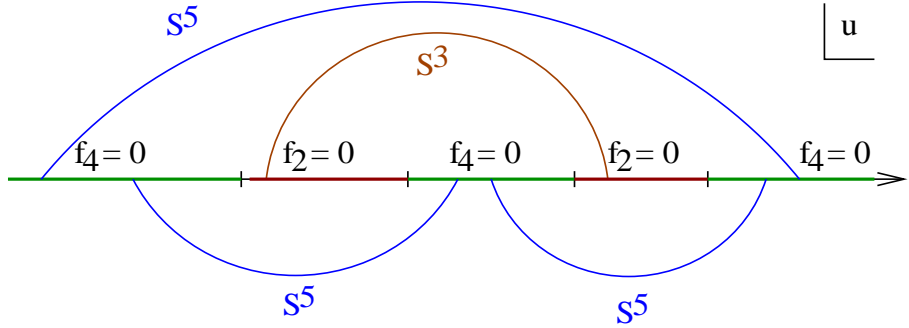}
\label{figure3}
\caption{The geometry of the genus 1 solution.}
\end{center}
\end{figure}

\subsection{Evaluation of the harmonic functions $h_1$ and $h_2$}

It is straightforward to calculate $h_2$, and we find,
\bea
\label{h2form}
h_2 = - 2i \left ( {1 \over \wp (z) - \wp (w_0)} - {1 \over \overline{\wp (z)} - \wp (w_0)} \right ) = { 2 i (\wp (z) - \wp (\bar z)) \over |\wp(z) - \wp (w_0)|^2}
\eea
To calculate $h_1$, we need the following integrals for $n=1,2$,
\bea
I_n (z,w_0) \equiv \int {dz \over (\wp (z) - \wp (w_0))^n}
\eea
These integrals may be calculated explicitly, using  the addition formula
for the $\zeta$-function,
\bea
\label{basic1}
\zeta (z+w_0) 
= \zeta (z) + \zeta (w_0) + \half {\wp '(z) - \wp '(w_0) \over \wp(z)-\wp(w_0)}
\eea
recast in the following form obtained by antisymmetrizing $w_0$,
\bea
\label{Widentity1}
\zeta (z+w_0) - \zeta (z-w_0) 
= 2 \zeta (w_0) - {\wp '(w_0) \over \wp (z) - \wp (w_0)}
\eea
Integrating the right hand side in $z$ yields the integral we need 
which is proportional to  $B_1$,
\bea
\label{ellint}
I_1(z,w_0)  = 
{1 \over \wp '(w_0)} \bigg ( \ln \sigma (z-w_0) - \ln \sigma (z + w_0)
+ 2 \zeta (w_0) z \bigg )
\eea
up to an additive integration constant. The integral we need proportional to
$B_2$ may be derived from (\ref{ellint}) by differentiating in $w_0$,
\bea
I_2 (z,w_0) & = &  {1 \over \wp '(w_0)^2} \bigg ( - \zeta (z-w_0) 
- \zeta (z+w_0) - 2 z \wp (w_0)
\no \\ && \hskip 1in 
- \half \left ( 12 \wp (w_0)^2 - g_2 \right ) I_1(z,w_0) \bigg )
\eea
where we have used the standard notation for the modular form 
defined by $g_2= 4 (e_1e_2+e_2e_3+e_3e_1)$.
Putting all together, we obtain the following expression for $h_1$, 
\bea
h_1 = B_0 - 2 i B_2 \bigg( I_2(z, w_0) - I_2(\bar z, w_0) \bigg) 
- 2 i B_1 \bigg( I_1(z,w_0) - I_1(\bar z,w_0) \bigg) 
- 2 i  \bigg( z - \bar z \bigg) \quad
\eea
where $B_0$ is a real integration constant.

\subsection{Vanishing Dirichlet boundary conditions}

In order to have a non-singular solution, the harmonic functions must vanish
on the Dirichlet parts of the boundary.  Attention only needs to be paid
to $h_1$ since $h_2$ always satisfies Dirichlet boundary conditions on the 
real $z$-axis. The boundary structure for $h_1$ is
\bea
\label{ellipticboundaryconditions}
-\infty < u < e_3: &\qquad& \p h_1 = \mbox{real} \qquad \mbox{Neumann}
\no\\
e_3 < u < e_2: &\qquad& \p h_1 = \mbox{imag} \qquad \mbox{Dirichlet}
\no\\
e_2 < u < e_1: &\qquad& \p h_1 = \mbox{real} \qquad \mbox{Neumann}
\no\\
e_1 < u < \infty: &\qquad& \p h_1 = \mbox{imag} \qquad \mbox{Dirichlet}
\eea
On the $z$ half-plane, the vanishing Dirichlet requirements are given by
\bea
h_1(\rho) = h_1(\rho + \omega_3) = 0 \hskip 1.0in \rho \in \mathbf{R}
\eea
Evaluating the harmonic function on the real axis gives
\bea
h_1 &=& B_0 - {2 i \over \wp^\prime(w_0)} 
\bigg( B_1 - {B_2 (12 \wp(w_0)^2 - g_2) \over 2 \wp^\prime(w_0)^2} \bigg)
\\ && \hskip 0.5in \times
\bigg( \ln  \sigma(z - w_0)  - \ln  \sigma( \bar z - w_0)  
- \ln  \sigma(z + w_0)  + \ln  \sigma( \bar z + w_0) \bigg)
\no
\eea
We leave in the log functions since discontinuities may arise from crossing 
branch cuts, all other functions are singly valued (when $z$ is restricted to 
region (I) in the fundamental domain).
To analyze the discontinuities, we first choose the branch cut for the log 
function to run along the negative real axis.  Consider first
the case when $\Re(z) > w$.  As $z$ approaches the real axis 
$\sigma(z \pm w_0)$ and $\sigma(\bar z \pm w_0)$ have real parts greater 
than zero and imaginary parts above or below the real axis.  Since
the branch cut runs along the negative axis the log terms simply cancel 
each other.  Requiring $h_1$ to vanish leads to the condition $B_0 = 0$.
Now consider the case when $\Re(z) < w_0$, for simplicity take $\Re(z) = 0$.  
Now $\sigma(z + w_0)$ and $\sigma(\bar z + w_0)$ again have real parts 
on the positive real axis, but $\sigma(z - w_0)$ and $\sigma(\bar z - w_0)$ 
have real parts along the negative real axis.  This is understood from the fact $\sigma(z)$ is an odd function of $z$.  Now as $z$ approaches the real axis, 
the log terms involving $z-w_0$ and $\bar z - w_0$ pick up a discontinuity 
from the branch cut and no longer cancel.  This is because the 
$\sigma(z - w_0)$ approaches the real axis from above while 
$\sigma(\bar z - w_0)$ approaches the real axis from below.  
This implies that their pre-factor must vanish which yields the condition
\bea
B_1 = {B_2 (12 \wp(w_0)^2 - g_2) \over 2 \wp^\prime(w_0)^2}
\eea
The above conditions are sufficient to ensure that $h_1$ vanishes along the 
real axis.  To enforce the vanishing of $h_1$ on the second Dirichlet boundary it suffices 
to require $h_1$ to vanish at $\omega_3$.  This gives $B_2$ as
\bea
1 = {B_2 \over \omega_3 \wp^\prime(w_0)^2} \bigg( 2 \omega_3 \wp(w_0) + \zeta(\omega_3 - w_0) + \zeta(\omega_3+ w_0) \bigg)
\eea
This relation may be simplified using the formula
$\zeta(\omega_3 - w_0) + \zeta(\omega_3 + w_0) = 2 \zeta(\omega_3)$, 
which can be derived from (\ref{Widentity1}) and the fact $\wp^\prime(\omega_i) = 0$.  
The Dirichlet constraints have now completely fixed the polynomial coefficients.  
There remains the two half-periods $\omega_i$ and the point
$u_0$.  We may fix $\omega_1 = 1$ by a scaling transformation, leaving $\omega_3$ and $u_0$ as the remaining parameters.  In summary, after satisfying vanishing Dirichlet boundary conditions, the harmonic functions are given by
\bea
\label{ellipticsolution}
h_1 &=& 2 i 
\bigg( 
 \zeta(z - w_0) + \zeta(z + w_0) - 2 {\zeta(\omega_3) \over \omega_3} z - {\rm c.c.} \bigg)
\no\\
h_2 &=& +{i \over \wp^\prime(w_0)} \bigg( 
\zeta(z + w_0) - \zeta(z - w_0) - {\rm c.c.}
\bigg)
\eea
where we have dropped an overall positive coefficient of 
$ [2  \wp(w_0) + 2 \zeta(\omega_3)/\omega_3 ]^{-1}$
in the expression for $h_1$.  By transformations $1.$ and $2.$ of subsection 7.7, 
this results in a constant shift of the dilaton, and an overall constant re-scaling 
of the metric and fluxes.  In the expression for $h_2$ we have made use of the 
identity $(\ref{Widentity1})$ to rewrite the inverse of the Weierstrass function in
terms of the Weierstrass zeta function.

\subsection{Positivity conditions and regularity}

The harmonic function $h_2$ is manifestly positive in view of the 
second relation in (\ref{h2form}).  In order to study the positivity of $h_1$, 
it will be useful to have a series expansion for the Weierstrass 
$\zeta$- function  in terms of trigonometric functions,
\bea
\zeta(z) = z {\zeta(\omega_3) \over \omega_3} + {\pi \over 2 \omega_3} 
\sum_{m = - \infty}^\infty \cotg 
\bigg({\pi z \over 2 \omega_3} + {m \pi \omega_1 \over \omega_3} \bigg)
\eea
The harmonic function $h_1$ is then given by,
\bea
h_1 = { i \pi \over \omega _3}  
\sum_{m = - \infty}^\infty \bigg [
\cotg \left ( {\pi \over 2 \omega_3} (z + w_0 + 2m \omega _1) \right )
+ \cotg \left ( {\pi \over 2 \omega_3} (z - w_0 + 2 m \omega _1)  \right ) \bigg]
+ {\rm c.c. } \quad
\eea
This may be simplified using the trigonometric identity,
\bea
\cotg \, u + \cotg \, \bar u = {\sin(u + \bar u) \over |\sin(u)|^2}
\eea
and we obtain,
\bea
h_1 (z, \bar z) & = & { i \pi \over \omega_3} \, h(z, \bar z) \, 
\sin \left ( {\pi \over 2 \omega_3} (z - \bar z) \right ) 
\\
h (z, \bar z) & = & 
 \sum_{m = - \infty}^\infty \left  \{
{1 \over \bigg|\sin \bigg({\pi \over 2 \omega_3} (z + w_0 + 2 m \omega_1) \bigg)\bigg|^2 }
+  
{1 \over \bigg | \sin \bigg({\pi \over 2 \omega_3} (z - w_0 + 2 m \omega_1) \bigg)\bigg|^2 } 
\right \}
\no
\eea
The harmonic function $h_1$ is manifestly positive.  We have now satisfied the regularity
requirements (R1)-(R4) of section 9.  It sill remains to satisfy requirement (R5),
which is $N_1 W > 0$ throughout $\Sigma$ and vanishes only on $\p \Sigma$.
While we have not been able to show this analytically, numerical investigations, 
shown \footnote{In figure 4, a multiplicative factor of $|z-w_0|^{10}$ has been included to 
regularize the singularity at $z = w_0$ which corresponds to the asymptotic 
$AdS_5 \times S^5$ region, while in figures 5 and 6, an analogous  multiplicative factor of 
$|z - w_0|^2$ has been included.} 
in figure 4,
indicate that requirement (R5) holds in the elliptic case for any values of the parameters
$\omega_3$, $w_0$.  
\begin{figure}[tbph]
\begin{center}
\epsfxsize=6in
\epsfysize=4in
\epsffile{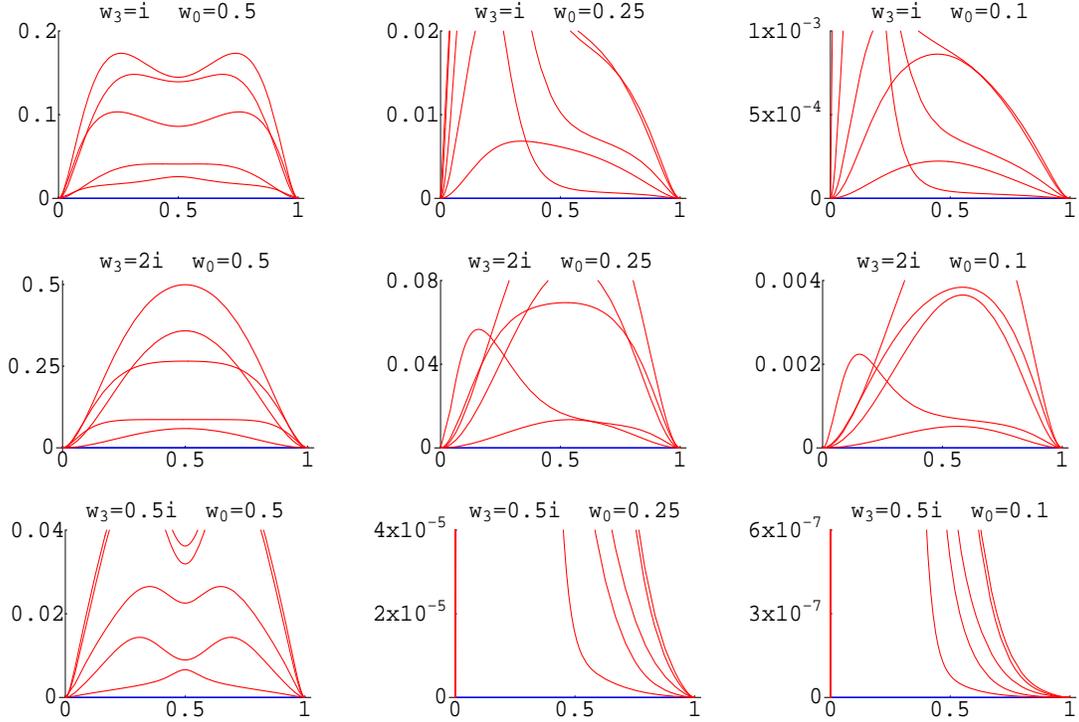}
\label{figure4}
\caption{Plots of $(- N_1 W)$ versus $\Re(z)$ for values of $\omega_3 = {i \over 2}, i, 2 i$ 
and $w_0 = {1 \over 2}, {1 \over 4}, {1 \over 10}$.  The different lines in each plot are for 
evenly spaced values of $\Im(z)$ with $0 < \Im(z) < \Im(\omega_3)$ (red) and 
$\Im(z) = 0, \Im(\omega_3)$ (blue).  On $\p \Sigma$, $(-N_1 W) = 0$ as expected. }
\end{center}
\end{figure}

\subsubsection{Explicit solution with $\omega_3 = i$ and $w_0 = \half$}

Here, we present plots of the dilaton and the metric factors in 
the case $\omega_3 = i$ and $w_0 = \half$.  It is clearly from
figure 5 that the dilaton is not constant, so that the solution is different
from $AdS_5 \times S^5$.  The metric factor $f_1$ never vanishes,
while $f_2$ and $f_4$ vanish only on $\p \Sigma$.  

\begin{figure}[tbph]
\begin{center}
\epsfxsize=6in
\epsfysize=2.1in
\epsffile{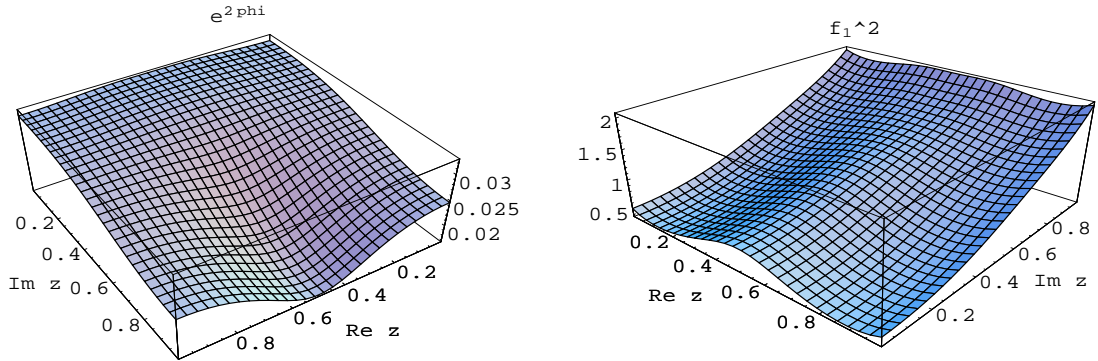}
\label{figure5}
\caption{Plots of the dilaton (left) and the metric factor $f_1^2$ (right) for the elliptic case
with $\omega_3 = i$ and $w_0 = \half$.  }
\end{center}
\end{figure}

\begin{figure}[tbph]
\begin{center}
\epsfxsize=6in
\epsfysize=2.1in
\epsffile{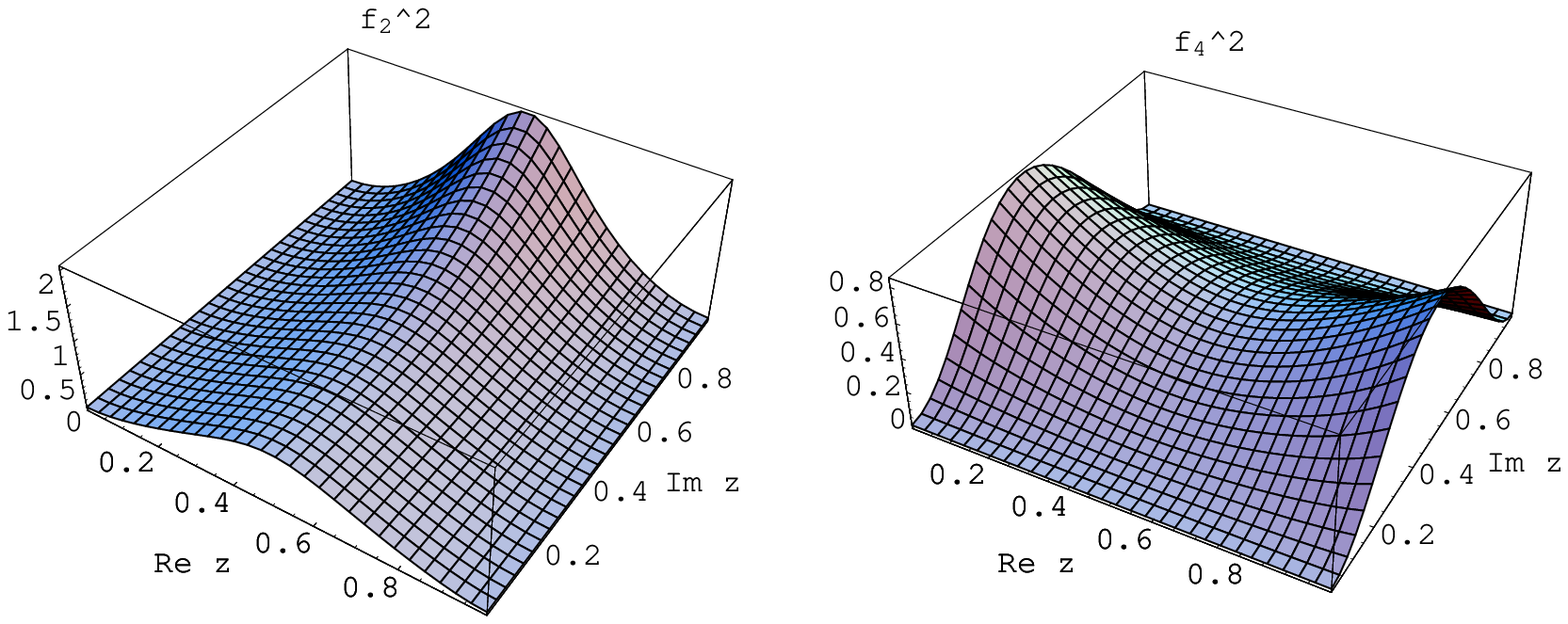}
\label{figure6}
\caption{Plots of the metric factor $f_2^2$ (left) and the metric factor $f_4^2$ 
(right) for the elliptic case with $\omega_3 = i$ and $w_0 = \half$.}
\end{center}
\end{figure}

\newpage

\section{Collapse of branch cuts}
\setcounter{equation}{0}

We recall the general ordering condition (\ref{order}) on the branch points and zeros
of the genus $g$ hyperelliptic solution, 
\bea
\a _{g+1} < e_{2g+1} <  \cdots < 
e_{2b+1} < e_{2b} < \a _b <  \cdots  < e_1 < u_0
\eea
for $b=1, \cdots , g$. From this condition,  it is clear that consecutive pairs of branch points 
come in two varieties, according to whether their firbration involves an $S^2$ or an $S^4$,
\bea
[e_{2b+1}, e_{2b} ] \times _f S^2 & \hskip 0.5in & {\rm D5-brane}
\no \\
{}[e_{2b}, e_{2b-1} ] \times _f S^4 && {\rm D3-brane}
\eea
Here, the product $\times _f$ stands for a fibration, not for a product of sets.
The collapse of a branch point in each variety corresponds to the collapse
of the corresponding $S^3$ or $S^5$, since the radii of $S^2$ and $S^4$ are always 1.

\smallskip

This situation is to be contrasted with the $AdS_4 \times S^2 \times S^2 \times \Sigma$
case in \cite{EDJEMG2}, where each segment between consecutive branch points always 
corresponded to a homology 3-sphere. The shrinking of one of these $S^3$ in  a genus
$g$ solution was regular or singular depending on how the limit of the real zeros
was taken. We showed in \cite{EDJEMG2} that a singular limit exists in
which naked D5-branes and NS5-branes are recovered. These naked solutions, 
in turn admitted a probe limit, which reflected the familiar singularity structure of 
the D5- and NS5-branes in flat space-time.

\smallskip

The situation for the $AdS_2 \times S^2 \times S^4 \times \Sigma$ solutions of this
paper is rather different. We shall show that the limit of collapsing branch cuts
of a regular genus $g$ hyperelliptic solution only leads to the non-singular solution
of genus $g-1$, but there is no room for naked D5-branes, NS5-branes, or D3-branes.
To keep calculations as explicit as possible, we illustrate this effect on the genus 1 solution.

\subsection{The genus 1 solution}

We study the collapse of a branch cut in the genus 1 case, where the ordering 
of branch points and zeros is given as follows, 
\bea
\alpha_2 < e_3 < e_2 < \alpha_1 < e_1 < u_0
\eea
The collapse of the segment $[e_3, e_2] $ corresponds to a collapsing $S^3$,
while the collapse of the segment $[e_2,e_1]$ (or of $[- \infty, e_3]$)
corresponds to a collapsing $S^5$. The behaviors of these two collapses 
are qualitatively different. We recall the expression for the differential $\p h_1$
and the harmonic function $h_2$ in the $u$-coordinates, 
\bea
\p h_1 & = & -i \, {(u-\a_1) (u- \a_2) du \over (u-u_0)^2 \sqrt{(u-e_1)(u-e_2)(u-e_3)}} 
\no \\
h_2 & = & -i \left ( { 1 \over u- u_0} - {1 \over \bar u - u_0} \right )
\eea
where the square root is taken to be negative for $u> e_1$.

\subsection{The collapse of $[e_2,e_1]$}

As $e_1-e_2 \to 0$, we necessarily also have $ \a _1 \to e_1=e_2$.
As a result, the differential $\p h_1$ simplifies considerably,
\bea
\p h_1 = -i \, {(u- \a_2) du \over (u-u_0)^2 \sqrt{u-e_3}} 
\eea
Comparison with (\ref{AdS5}) immediately reveals that this is just the 
$AdS_5 \times S^5$ solution at genus 0, with $\a_2 = e_3 - 1/2$.
In figure 1, this collapse corresponds to the upper $S^5$ shrinking to
0. Note that in the process, the homology 3-sphere disappears as well.

\subsection{The collapse of $[e_3,e_2]$}

By an overall translation of $u$, we may fix $e_1 = 0$, and it will be convenient to 
designate the shifted branch points by $k^2 \equiv - e_3 = - e_2$, with $k > 0$.  
The collapse of the branch points  for generic $\alpha_1$ and $\alpha_2$ would leave a 
pole located in a region with vanishing $S^2$.  But $\a_1$ and $\a_2$ are 
actually completely determined in terms of $u_0$ and the moduli. Here,
we shall recover these relations directly from the genus 1 solution.
To uniformize the square root, we introduce the coordinate $w^2 \equiv u$, 
with $w_0^2 \equiv u_0$, and $w$ takes values in the second quadrant, 
$\Re(w)<0$ and $\Im(w) > 0$.  
In these coordinates, the differential becomes simply
\bea
\p h_1 = - 2 i {(w^2 - \alpha_1) (w^2 - \alpha_2) \over (w^2 - w_0^2)^2 (w^2 + k^2) } dw
\eea
To integrate we decompose in partial fractions
\bea
\p h_1 = - 2 i \bigg( 
{A \over (w - w_0)^2} + {A \over (w + w_0)^2} 
+ {B \over w - w_0} - {B \over w + w_0}
+ {i C \over w + i k} - {i C \over w - i k}
\bigg) dw
\eea
where the constant coefficients are given by
\bea
\label{polecoefficients}
&& A = {(w_0^2 - \alpha_1) (w_0^2 - \alpha_2) \over 4 w_0^2 (w_0^2 + k^2)} > 0
\no\\
&& B = {1 \over 2 w_0} \left ( 1 - 2 A-2 k C \right )
\no\\
&& C = {(k^2 + \alpha_1) (k^2 + \alpha_2) \over 2 k (w_0^2 + k^2)^2} < 0
\eea
Integrating the harmonic function, we obtain
\bea
\label{poleharmonicfunction}
&& h_1 = - 2 i \bigg[ D 
- A \bigg({1 \over w - w_0} + {1 \over w + w_0} \bigg)
+ B \ln \bigg( {w - w_0 \over w + w_0} \bigg)
- i C \ln \bigg( {w - i k \over w + i k} \bigg) - {\rm c.c.} \bigg]
\no\\
&&
h_2 = - i \bigg[ {1 \over w^2 - w_0^2} - {\rm c.c.} \bigg]
\eea
where $D$ is a purely imaginary integration constant.

\smallskip

The boundary structure is the same as in the elliptic case (\ref{ellipticboundaryconditions}), 
but with the branch points $e_2$ and $e_3$ identified.  Translating to the $w$-plane, 
$h_1$ satisfies  Neumann conditions along the imaginary $w$-axis, and vanishing Dirichlet conditionas along the real $w$-axis.
In these coordinates, real infinity and imaginary infinity correspond to the same branch point
in the $u$-plane, so $h_1$ should also be required to vanish at imaginary infinity. 
The explicit conditions are
\bea
\label{poledirichlet}
h_1(\rho) = 0 \qquad \rho \in \bR
\qquad \qquad
h_1(i \rho) = 0 \qquad \rho \rightarrow \infty
\eea
Vanishing at $w=\infty$ clearly requires $D=0$. Vanishing on the left and right 
of $w_0$ requires that the coefficient of the branch cut discontinuity of $\ln (w-w_0)$
vanish as well, so that we must have $B=0$. Once $B=D=0$, it is manifest that 
$h_1$ vanishes at imaginary infinity,
and that all of the vanishing Dirichlet conditions (\ref{poledirichlet}) are satisfied.

\smallskip

The term proportional to $A>0$ in $h_1$ is easily seen to be positive for $w$
in the second quadrant. But the term proportional to $C<0$ is negative there.
In particular, as $w \to i k$, this term would make $h_1$ tend to $-\infty$,
producing a behavior which is unacceptable for a physical solution.
Thus, we must have $C=0$, which in turn require that $ \a_1=-k^2$ or
$\a_2 =-k^2$. The vanishing of $B$ now requires that $2A=1$, which 
determined $\a_2 = - w_0^2$ or $\a_1= - w_0^2$ respectively. In either
case do we recover the genus 0 solution which is $AdS_5 \times S^5$.

\bigskip

\bigskip

\noindent{\bf Acknowledgments}

\medskip 

\noindent This work was supported in
part by National Science Foundation (NSF) Physics Division grant PHY-04-56200.

\newpage

\appendix

\section{Clifford algebra basis adapted to the Ansatz}
\setcounter{equation}{0}
\label{appA}

We use the convention $\eta = {\rm diag} [-+\cdots  + ]$, and
choose a basis for the Clifford algebra which is well-adapted to the
$AdS_2 \times S^2 \times  S^4 \times  \Sigma$ Ansatz, with the frame
labeled as in (\ref{frame1}),
\bea
\G^\mu & = & \g^\mu \, \otimes \,  I_2 \, \, \otimes \, I_4 \, \otimes \, I_2 \hskip 1.1in \mu=0,1
\no \\
\G^i & = & \g_{(1)} \otimes \g^i \, \otimes \, I_4 \, \otimes \, I_2 \hskip 1.1in i = 2,3
\no \\
\G^m & = & \g_{(1)} \otimes \g_{(2)} \otimes \g^m \otimes I_2 \hskip 1in m =4,5,6,7
\no \\
\G^a & = & \g_{(1)}  \otimes \g_{(2)} \otimes \g_{(3)} \otimes \gamma ^a \hskip 1in a=8,9
\eea
where  a convenient basis for the lower-dimensional Clifford algebras is as follows,
\bea
i \g^0 = \sigma ^2 & \hskip 1in & \g^4 = \sigma ^1 \otimes I_2
\hskip 1.2in \g^8 = \sigma ^1
\no \\
\g^1 = \sigma ^1  & \hskip 1in & \g^5 = \sigma ^2 \otimes I_2
\hskip 1.2in \g^9 = \sigma ^2
\no \\
\g^2 =  \sigma ^2 & \hskip 1in & \g^6 = \sigma ^3 \otimes \sigma ^1
\no \\
\g^3 =  \sigma ^1& \hskip 1in & \g^7 = \sigma ^3 \otimes \sigma ^2
\eea
so that the chirality matrices take the form,
\bea
\g_{(1)} = - \g^{01} = \sigma ^3 & \hskip 1in &
    \g_{(3)}  = - \g ^{4567} = \sigma ^3 \otimes \sigma ^3
\no \\
\g_{(2)} =\, i \g^{23} = \sigma ^3 & \hskip 1in &
\gamma _{(4)} = - i \g ^{89} =  \sigma ^3
\eea
We shall also need the chirality matrices on the various components of
$AdS_2 \times S^2 \times S^4 \times   \Sigma$, and they are chosen as follows,
\bea
\G _{(1)} & = & - \G^{01} ~ ~ \, = \g_{(1)} \otimes  I_2 \otimes I_4 \otimes I_2
\no \\
\G _{(2)} & = & +i \G^{23} ~~ = I_2 \otimes \g_{(2)} \otimes  I_4 \otimes I_2
\no \\
\G _{(3)} & = & - \G^{4567} ~ = I_2 \otimes I_2 \otimes \g_{(3)}  \otimes I_2
\no \\
\G _{(4)} & = & -i \G^{89} ~~ = I_2 \otimes I_2 \otimes I_4 \otimes \g_{(4)}
\eea
The 10-dimensional chirality matrix in this basis is given by
\bea
\G^{11} = \G^{0123456789}   = \G_{(1)} \G_{(2)} \G_{(3)} \G_{(4)}
= \g_{(1)} \otimes \g_{(2)} \otimes \g_{(3)} \otimes \g_{(4)}
\eea
The complex conjugation matrices in each component are defined by
\bea
\left ( \g^\mu \right ) ^* = + B_{(1)} \g ^\mu B_{(1)} ^{-1}
& &   (B_{(1)})^* B_{(1)} = + I_2 \hskip .6in B_{(1)} =  I_2
\no \\
\left ( \g^i \right ) ^* = - B_{(2)} \g ^i B_{(2)} ^{-1}
& &   (B_{(2)})^* B_{(2)} = - I_2 \hskip .6in B_{(2)} =  \g^2 = \sigma ^2
\no \\
\left ( \g^m \right ) ^* = - B_{(3)} \g ^m B_{(3)} ^{-1}
& \hskip .5in &   (B_{(3)})^* B_{(3)} = - I_4 \hskip .6in B_{(3)}
= i \g^{46} = \sigma ^2 \otimes \sigma ^1
\no \\
\left ( \g^a \right ) ^* = - B_{(4)} \g ^a B_{(4)} ^{-1}
& &   (B_{(4)})^* B_{(4)} = - I_2 \hskip .6in B_{(4)} =  \g^9 = \sigma ^2
\eea
where in the last column we have also listed the form of these matrices
in our particular basis. It is also useful to note how the $B_{(i)}$ commute
past the chirality matrices $\g_{(i)}$
\bea
B_{(1)} \g_{(1)} & = & + \g_{(1)} B_{(1)}
\no \\
B_{(2)} \g_{(2)} & = & - \g_{(2)} B_{(2)}
\no\\
B_{(3)} \g_{(3)} & = & + \g_{(3)} B_{(3)}
\no \\
B_{(4)} \g_{(4)} & = & - \g_{(4)} B_{(4)}
\eea
The 10-dimensional complex conjugation matrix $\cB$
is defined by $(\G^M)^* = \cB \G^M \cB^{-1} $ and $\cB \cB^* = I$, and in this
basis is given by
\bea
\label{10dconjugationmarix}
\cB =   - \G^{2579}
    & = & + i B_{(1)} \otimes \g_{(2)} B_{(2)} \otimes
     B_{(3)} \otimes  B_{(4)}
    \no \\
& = &  I_2 \otimes \sigma ^1 \otimes \sigma ^2 \otimes \sigma ^1 \otimes \sigma ^2
\eea
and satisfies $\cB^* = \cB$, and $\cB^2 = I$.

\smallskip

It is useful to check that the self-duality of $F_{(5)}$ is compatible with the
chirality conventions of the $\G$-matrices. To this end, we evaluate
\bea
\G^{0123} = i \G_{(1)} \G_{(2)}
& \hskip .7in &
{1 \over 5!} (F_{(5)} \cdot \G) = - i f_a \G_{(1)} \G_{(2)} \G^a \left ( I + \G^{11} \right )
\no \\
\G^{4567} = - \G_{(3)} \hskip .2in
&  &
\eea
which has the correct projection properties. Here, we have used,
$\ep^{ab} \g^b = i \g^a \g_{(4)}$, $a,b=8,9$.

\newpage

\section{The geometry of Killing spinors}
\setcounter{equation}{0}
\label{appB}

We review the relation between the Killing spinor
equation and the parallel transport equation in the presence of a flat connection
with torsion on $AdS_2$, and even-dimensional spheres.

\subsection{Minkowski $AdS_2$}

For Minkowski signature, we have $AdS_2 = SO(2,1)/SO(1,1)$.
The Clifford algebra of $SO(2,1)$ is built from the Clifford generators
$\gamma ^\mu$,  with $\mu =0,1$,
\bea
\{ \gamma ^\mu , \gamma ^\nu\} = 2 \eta ^{\mu \nu }
\hskip 1in
\eta = {\rm diag} [- +]
\eea
and the matrix, $\gamma ^{\sharp } \equiv i \g^0 \g^1 $,
which is proportional to the chirality matrix, but has  $(\gamma^\sharp)^2 = -1$.
The matrices $\g^\sharp, \g^0, \g^1$ form a 2-dimensional representation of the
Clifford algebra,
\bea
\{ \gamma ^ {\bar \mu} , \gamma ^{\bar \nu} \} = 2 \bar \eta ^{\bar \mu \bar \nu}
 \hskip 1in
\bar \eta = {\rm diag} [- - +]
\eea
for $\bar \mu, \bar \nu = \sharp, 0,1$.
The corresponding Maurer-Cartan form on $SO(2,1)$ is given by
\bea
\omega ^{(t)} = V ^{-1} dV = {1 \over 4} \omega ^{(t)} _{\bar \mu \bar \nu}
\gamma ^{\bar \mu \bar \nu} \hskip 1in V \in SO(2,1)
\eea
It obviously satisfies the Maurer-Cartan equations, $d \omega ^{(t)}
+ \omega ^{(t)} \wedge \omega ^{(t)}=0$.
We decompose $\omega ^{(t)}$ onto the $SO(1,1)$ connection and cotangent space
frame on $AdS_2$
\bea
\omega ^{(t)} = { 1 \over 4} \omega _{\mu \nu} \gamma ^{\mu \nu}
+ {1 \over 2} e_\mu \gamma ^\mu \gamma ^{\sharp}
\hskip 1in
\left \{ \matrix{
\omega _{\mu \nu}   \equiv \omega ^{(t)} _{\mu \nu } & \mu,\nu =0,1 \cr
e_\mu \equiv \omega ^{(t)} _{\mu \sharp} & \mu=0,1 \cr} \right .
\eea
The Maurer-Cartan equations $ d \omega ^{(t)} + \omega ^{(t)} \wedge \omega ^{(t)}=0$
for $\omega ^{(t)}$ imply the absence of torsion and the constancy of curvature.
The Killing spinor equation coincides with the equation for parallel transport,
\bea
\left ( d + V^{-1} dV \right ) \ep \left ( d + { 1 \over 4} \omega _{\mu \nu} \gamma ^{\mu \nu}
+ {1 \over 2} \eta e_\mu \gamma ^\mu \gamma^{\sharp}  \right ) \ep =0
\eea
For $\eta=+1$, the general solution is given by $\ep _+ = V^{-1} \ep _0$
and $\ep_0$ is constant, while for $\eta =-1$, the solution is $\ep_- = \gamma ^\sharp \ep _+$.

\subsection{Even-dimensional spheres}

The sphere is a coset, $S^N = SO(N+1)/SO(N)$, and we consider the case $N$ even.
The Clifford algebra of $SO(N+1)$ is built from the Clifford generators
$\gamma ^m$  of  $SO(N)$,
\bea
\{ \gamma ^m , \gamma ^n \} = 2 \delta ^{mn }
\eea
for $m,n =1,2,...,N$, supplemented with the chirality matrix,
$\gamma ^{\sharp } \equiv (i) \gamma ^{1...N} $,
where the $(i)$ is chosen so that $(\gamma ^{\sharp})^2 = + 1$,
\bea
\{ \gamma ^ {\bar m} , \gamma ^{\bar n} \} = 2 \delta ^{\bar m \bar n}
\eea
for $\bar m, \bar n = \sharp , 1,2,...,N$.
The corresponding Maurer-Cartan form on $SO(N+1)$ is given by
\bea
\omega ^{(t)} = V ^{-1} dV = {1 \over 4} \omega ^{(t)} _{\bar m \bar n}
\gamma ^{\bar m \bar n} \hskip 1in V \in SO(N+1)
\eea
It obviously satisfies the Maurer-Cartan equations, $d \omega ^{(t)}
+ \omega ^{(t)} \wedge \omega ^{(t)}=0$.
We decompose $\omega ^{(t)}$ onto the $SO(N)$ and $S^N$
directions of cotangent space,
\bea
\omega ^{(t)} = { 1 \over 4} \omega _{mn} \gamma ^{mn}
+ {1 \over 2} e_m \gamma ^m \gamma ^{\sharp}
\hskip 1in
\left \{ \matrix{
\omega _{mn}   \equiv \omega ^{(t)} _{mn } & m,n =1,2, \cdots ,  N \cr
e_m \equiv \omega ^{(t)} _{m \sharp} & m=1,2, \cdots , N \cr} \right .
\eea
The Maurer-Cartan equations $ d \omega ^{(t)} + \omega ^{(t)} \wedge \omega ^{(t)}=0$
for $\omega ^{(t)}$ imply the absence of torsion and the constancy of curvature.
The Killing spinor equation coincides with the equation for parallel transport,
\bea
\left ( d + V^{-1} dV \right ) \ep= \left ( d + { 1 \over 4} \omega _{mn} \gamma ^{mn}
- {1 \over 2} \eta e_m \gamma ^m \gamma ^{\sharp}  \right ) \ep =0
\eea
For $\eta=-1$, the general solution is given by $\ep _- = V^{-1} \ep _0$
and $\ep_0$ is constant, while for $\eta =+1$, the solution is $\ep_+ = \gamma ^\sharp \ep _-$.

\subsection{Explicit form of Killing equations on  $AdS_2,$ $S^2,$ and $S^4$}

$\bullet$
For $AdS_2$ we have $\g^0=-i \sigma ^2$, $\g^1 = \sigma ^1$, and
$\gamma ^\sharp= i \gamma _{(1)} = i \sigma ^3$. The Killing spinor equation is then,
for $\mu = 0,1$,
\bea
\bigg( \hat \nabla_\mu - i \, {\eta_1 \over 2} \gamma_\mu  \gamma_{(1)} \bigg) \ep = 0
& \qquad \Leftrightarrow \qquad &
\bigg( \hat \nabla_\mu - {\eta_1 \over 2} \gamma_\mu  \bigg) \ep^\prime = 0
\eea
where $\ep^\prime = e^{- i {\pi \over 4} \gamma_{(1)}} \ep$, and $\eta _1 = \pm 1$.

\smallskip

\noindent $\bullet$
For  $S_2$ we have $\g^2=\sigma ^2$, $\g^3 = \sigma ^1$, and $\g^\sharp = \gamma _{(1)}$.
The Killing spinor equation is then, for $i=2,3$,
\bea
\bigg( \hat \nabla_i - {\eta_2 \over 2} \gamma_i  \gamma_{(2)} \bigg) \ep = 0
& \qquad \Leftrightarrow \qquad &
\bigg( \hat \nabla_i - i \, {\eta_2 \over 2} \gamma_i  \bigg) \ep^\prime = 0
\eea
where $\ep^\prime = e^{- i {\pi \over 4} \gamma_{(2)}} \ep$, and $\eta _2 = \pm 1$.

\smallskip

\noindent $\bullet$
For $S^4$, the $\g^m$, $m=4,5,6,7$ were given above and
$\g^\sharp = \g^{4567}= -\gamma _{(3)}$.
The Killing spinor equation is given by, for $m=4,5,6,7$,
\bea
\bigg( \hat \nabla_m - {\eta_3 \over 2} \gamma_m  \gamma_{(3)} \bigg) \ep = 0
& \qquad \Leftrightarrow \qquad &
\bigg( \hat \nabla_m - i \, {\eta_3 \over 2} \gamma_m  \bigg) \ep^\prime = 0
\eea
where $\ep^\prime = e^{- i {\pi \over 4} \gamma_{(3)} } \ep$, and $\eta _3 = \pm 1$.

\newpage

\section{The Reduced  Bianchi identities and Field Equations}
\setcounter{equation}{0}

Using the differential form notation of \S 1.2, it is straightforward to reduced the Bianchi
identities to this Ansatz. We find,
\bea
\label{redBianchi}
0 & = &  d P - 2 i Q \wedge P
\no \\
0 & = &  dQ + i P \wedge \bar P
\no \\
0 & = &  d\cG + 2 (d \ln f_1) \wedge \cG  -i Q \wedge \cG + P \wedge \bar \cG
\no \\
0  & = & d\cH + 2 (d \ln f_2) \wedge \cH  -i Q \wedge \cH - P \wedge \bar \cH
\no \\
0 & = &  d  (*_2 \cF) + 4 (d \ln f_4) \wedge (*_2 \cF )
\no \\
0 & = &  d \cF + 2 \left ( d \ln (f_1 f_2) \right ) \wedge \cF
+ {1 \over 8} \left ( \cG \wedge \bar \cH  + \bar \cG \wedge \cH \right )
\eea
The field equations of Type IIB supergravity, reduced to the two-parameter 
Ansatz of \S 1.2 are given as follows. 
The BPS equations will imply that every solution may be mapped to
one with vanishing axion and thus $Q=0$, and $g_a, h_a$ and $p_a$ real.

\smallskip

Using the convention $f^2 d B = d\phi$, the dilaton equation becomes,
\bea
D^a D_a \phi + 2 (D^a \phi) D_a \ln (f_1f_2 f_4)
- {1 \over 4} (g_a g^a + h_a h^a) =0
\eea
The 3-form field equations reduce to the following two real equations,
\bea
D^a g_a + 2 g_a D^a \ln (f_2 f_4^2)
- p^a  g_a + 4 f_a  h_a & = &0
\no \\
D^a h_a + 2 h_a D^a \ln (f_1 f_4^2)
+ p^a  h_a + 4 f_a  g_a & = &0
\eea
Finally, the Einstein equations, respectively for the components $mn$,
$i_1j_1$, $i_2j_2$, and $ab$ are as follows, (all other components must vanish by
$SO(2,3) \times SO(3) \times SO(3)$ symmetry),
\bea
\label{Einstein}
0 & = &
+ {3 \over f_4^2}
- 3 { |D_a f_4 |^2 \over f_4 ^2}
- 2 {D^a f_4 D_a (f_1 f_2) \over f_1 f_2 f_4}
- {D^a D_a f_4  \over f_4}
- 4 f_a f^a - {1 \over 8} g_a  g^a + {1 \over 8} h_a  h^a
\no \\ && \no \\
0 & = &
- {1 \over f_1^2}
-  { |D_a f_1 |^2 \over f_1 ^2}
- 4 {D^a f_4 D_a f_1 \over f_1  f_4}
- 2 {D^a f_1 D_a f_2 \over f_1  f_2}
- {D^a D_a f_1 \over f_1}
+ 4 f_a f^a + {3 \over 8} g_a  g^a + {1 \over 8} h_a  h^a
\no \\ && \no \\
0 & = &
{1 \over f_2^2}
-  { |D_a f_2 |^2 \over f_2 ^2}
- 4 {D^a f_4 D_a f_2 \over f_2  f_4}
- 2 {D^a f_1 D_a f_2 \over f_1  f_2}
- {D^a D_a f_2 \over f_2}
+ 4 f_a f^a - {1 \over 8} g_a  g^a - {3 \over 8} h_a  h^a
\no \\ && \no \\
0 & = &
- 4 {D_b D_a f_4 \over f_4}
- 2{D_b D_a f_1 \over f_1}
- 2{D_b D_a f_2 \over f_2}  + R^{(2)} \delta _{ab}
- 2 D_a \phi D_b \phi
- 4 \delta _{ab} f_c f^c + 8 f_a f_b
\no \\ && \hskip .3in
- {1 \over 8} \delta _{ab} \left ( g_c  g^c - h_c  h^c \right )
- {1 \over 2} \left (- g_a  g_b  +  h_a  h_b  \right )
\eea

\end{document}